\documentclass[10pt]{iopart}

\usepackage{iopams}
\usepackage{graphicx}
\newcommand{\SiOTwo}{SiO$_{\mathrm{2}}$}
\newcommand{\Si}{Si}
\begin{document}

\title{Kinetic modeling of the electronic response of a dielectric plasma-facing solid}

\author{Franz X. Bronold and Holger Fehske}

\address{Institut f\"ur Physik, Ernst-Moritz-Arndt-Universit{\"a}t Greifswald,
 17489 Greifswald, Germany}
\ead{bronold@physik.uni-greifswald.de}
\vspace{10pt}
\date{\today}

\begin{abstract}
We present a self-consistent kinetic theory for the electronic response of a plasma-facing
dielectric solid. Based on the Poisson equation and two sets of 
spatially separated Boltzmann equations, one for electrons and ions in the plasma and one 
for conduction band electrons and valence band holes in the dielectric, the approach gives 
the quasi-stationary density and potential profiles of the electric double layer forming at 
the interface due to the permanent influx of electrons and ions from the plasma. The two 
sets of Boltzmann equations are connected by quantum-mechanical matching conditions for the 
electron distribution functions and a semi-empirical model for hole injection mimicking the 
neutralization of ions at the surface. Essential for the kinetic modeling is the ambipolarity 
inside the wall, leading to an electron-hole recombination condition, and the merging of the 
double layer with the quasi-neutral, field-free regions deep inside the wall and the plasma.
To indicate the feasibility as well as the potential of the approach we apply it to a 
collisionless, perfectly absorbing interface using intrinsic and extrinsic silicon dioxide
and silicon surfaces in contact with a two-temperature hydrogen plasma as an example.
\end{abstract}


%
%
%
\ioptwocol

\section{Introduction}

The basic electronic response of a plasma-facing solid (wall) is the accumulation of 
electrons from the plasma. It acquires a negative charge because impacting electrons are 
deposited inside the solid more efficiently than electrons are extracted from it by the 
neutralization of positive ions. The negatively charged wall in turn triggers a 
positively charged space charge region in front of it: the plasma sheath~\cite{Franklin76,LL05}. 
The total result of the electronic response of the plasma-wall interface is thus an 
electric double layer. Yet, ever since the early days of gaseous electronics~\cite{LM24}, 
the negative part of the double layer--the wall charge--plays no essential role in the 
studies of the plasma sheath. Typically the focus is either on how the sheath merges with 
the quasi-neutral bulk plasma (see the reviews~\cite{SB90,Riemann91,Franklin03,Brinkmann09,Robertson13} 
and references therein) or on how the emissive properties of the wall, most notably, 
electron reflection (or absorption) and secondary electron emission, affect 
the spatial structure of the sheath~\cite{HZ66,TLC04,SKR09,GK12,Campanell15}. 

Evidently for this type of studies the wall is considered as a reservoir characterized 
by a geometrical boundary and probabilities for particle reflection, absorption, 
and emission which may be chosen ad-hoc to make plasma simulations
reproduce experimental findings, deduced from quantum-mechanical calculations~\cite{MBF12,BF15}, 
or--in very rare occasions--obtained from independent measurements~\cite{PP99,DAK15}. In 
situations where the length and time scales of the plasma and the wall are well separated 
this is a viable strategy. Even microdischarges~\cite{SchoenbachBecker16} with a linear 
extension of $15-40\,\mu {\rm m}$ can be successfully modeled by such an approach~\cite{Kushner04}. 
However, if this is not the case, or if the electrons accumulated in the solid are an integral 
part of the physical system one is interested in, as it is, for instance, the case for the plasma 
bipolar junction transistor~\cite{TWH11}, this approach is not sufficient. A kinetic description 
has then to be set up also for the wall and merged to the one of the plasma by suitable 
matching conditions. 

In particular, solid-state based integrated microdischarges~\cite{OE05,DOL10,KSO12,EPC13} 
can be expected to soon reach the sub-micron range where the electron transit time 
$\tau_e^{\rm transit}$ through the sheath of the discharge approaches the electron energy 
relaxation time $\tau_*^{\rm relax}$ inside the solid. In this case, the electronic subsystem 
of the wall remains out-of-equilibrium between subsequent electron encounters from the plasma 
and surface parameters have to be obtained for a solid in strong electronic non-equilibrium. 
Taking, as an illustration, a $1\,\mu{\rm m}$ wide microdischarge with a screening length 
$\lambda_D^p$ of one-tenth of its width, which seems to be feasible~\cite{EPC13}, 
and an electron temperature $k_BT_e=2\,{\rm eV}$ yields 
$\tau_e^{\rm transit}\approx\lambda_D^p/\bar{v}_e\approx 10^{-13}\,{\rm s}$, where
$\bar{v}_e=\sqrt{8k_BT/\pi m_e}$ is the thermal velocity of the electron. This is 
only one order of magnitude smaller than the typical electron energy relaxation time 
in the conduction band of a dielectric solid such as \SiOTwo , 
$\tau_*^{\rm relax}\approx l_{\rm inel}/v_*\approx 10^{-14}\,{\rm s}$, assuming 
a kinetic energy of the injected electron of $2\,{\rm eV}$ above the bottom of the 
conduction band, an effective mass of $0.8\,m_e$, and an inelastic scattering length 
of $l_{\rm inel}\approx 100\,\AA$ (from the universal curve~\cite{Lueth15}), but 
already much shorter than the electron-hole recombination time which is on the order 
of nanoseconds. In addition, the separation between integrated microdischarges can be 
made small enough to enable crosstalk through the space charge layers inside the 
wafer opening thereby perhaps opportunities for novel opto-electronic plasma 
devices. There are thus situations conceivable where the electronic processes in 
the solid and the plasma cannot be considered independently anymore. Focusing on the 
positive part of the electric double layer alone will then be of course also no 
longer sufficient.  

Double layers are abundant in nature and have been studied in various
contexts. They arise at any interface separating
positive and negative charges. In solid state physics the most important 
double layer is the pn-junction~\cite{Li06} which is at the heart of many 
electronic devices. Double layers occur also when two different gaseous plasmas
approach each other~\cite{AA71,SB83,Raadu89,Charles07}.  
The double layer at the plasma-wall interface is however special.  
On the one hand, and in contrast to pn-junctions, it is always far from equilibrium, 
involving very hot charge carriers. On the other hand, and this distinguishes it from 
the gaseous double layers, it is spatially pinned by the crystallographic ending of 
the wall and energetically constrained by the wall's band structure. Irrespective 
of the demands arising from the miniaturization of microdischarges studying 
double layers at plasma-wall interfaces is thus also of fundamental interest.

Up to now electric double layers at plasma-facing solids have not been studied
extensively. To the best of our knowledge metallic surfaces have not been investigated 
at all and there are only a few studies~\cite{TD85,HBF12} devoted to dielectric surfaces 
in contact with a plasma. But even for them a satisfactory description is still missing. 
In our own work~\cite{HBF12} on the subject, for instance, we 
employed a thermodynamical principle to distribute the electrons missing in 
the plasma sheath in a graded potential which interpolates between the sheath 
and the potential inside the wall~\cite{HBF12}. The basic assumption, however, 
that at quasi-stationarity the electrons are thermalized within the wall, is 
only valid for some of the electrons and not for all. In addition, the approach 
was based on drift-diffusion equations. Hence, the 
dynamical variables were from the start particle densities, fluxes and electric 
potentials. Distribution functions did not appear. The coupling of the positive 
and negative parts of the double layer was thus entirely due to the matching 
conditions for the electric potential. It was hence impossible to include
quantum-mechanical reflection of electrons by the surface potential and/or 
electron extraction due to the neutralization of ions.

Below we present a kinetic approach for a dielectric surface which is general enough 
to overcome these shortcomings. It works with the Poisson equation and two sets of 
Boltzmann equations operating in disjunct half-spaces. The matching at the interface 
is performed not only for the electric potential but also for the distribution 
functions which enables us to keep the ambipolarity of the plasma side (electrons 
and ions) alive inside the wall (electrons and holes). 
 Eventually, this allows us to formulate a recombination condition for electrons
   and holes, which in turn limits, in conjunction with further conditions imposed
   at quasi-stationarity, the continuous influx of electrons and ions from the
   plasma.
For the electron distribution functions the 
matching conditions are essentially identical with the matching conditions used for 
solid interfaces~\cite{DLP95,Falkovsky83,Schroeder92}. The matching condition for the
ion distribution function on the other hand is a semi-empirical model for electron 
extraction (that is, hole injection) connecting the ion distribution 
function of the plasma with the hole distribution function of the wall.
A thermodynamical principle is no longer used. Instead, we only demand as 
boundary conditions quasi-neutral, field-free regions far away from the interface. 

In the numerical calculations we consider a collisionless perfectly absorbing surface. 
But the kinetic approach is first described in broader terms so that it becomes clear 
how to include collisions and how to include quantum-mechanical reflection of electrons. 
The numerical solution of the complete kinetic model is however rather demanding
and beyond the scope of the present work. The numerical calculations yield thus 
only the self-consistent quasi-stationary potential and density profiles across 
an idealized interface. For given temperature and mass ratios the incoming flux
of electrons and ions is self-consistently determined. Despite the simplicity
of the model, the numerical results for \SiOTwo\ and \Si\ surfaces in contact 
with a hydrogen plasma are very promising and clearly indicate the feasibility and 
potential of our approach for revealing the rich physics taking place inside the 
wall.
 
The outline of the remaining part of the paper is as follows. In Sect.~\ref{Formalism} 
we describe in detail our approach, first, in general terms and then for the special
case of a collisionless, perfectly absorbing dielectric surface. Section~\ref{Results}
discusses representative numerical data for idealized intrinsic and doped \SiOTwo\ 
and \Si\ surfaces exposed to a two-temperature hydrogen plasma focusing on quasi-stationary
density and potential profiles. A discussion of the issues which need to be resolved 
before the approach becomes quantitative for realistic surfaces is given in the 
concluding Section~\ref{Conclusions}. Mathematical details interrupting the flow of the
discussion are relegated to an Appendix.

\begin{figure}[t]
\includegraphics[width=0.99\linewidth]{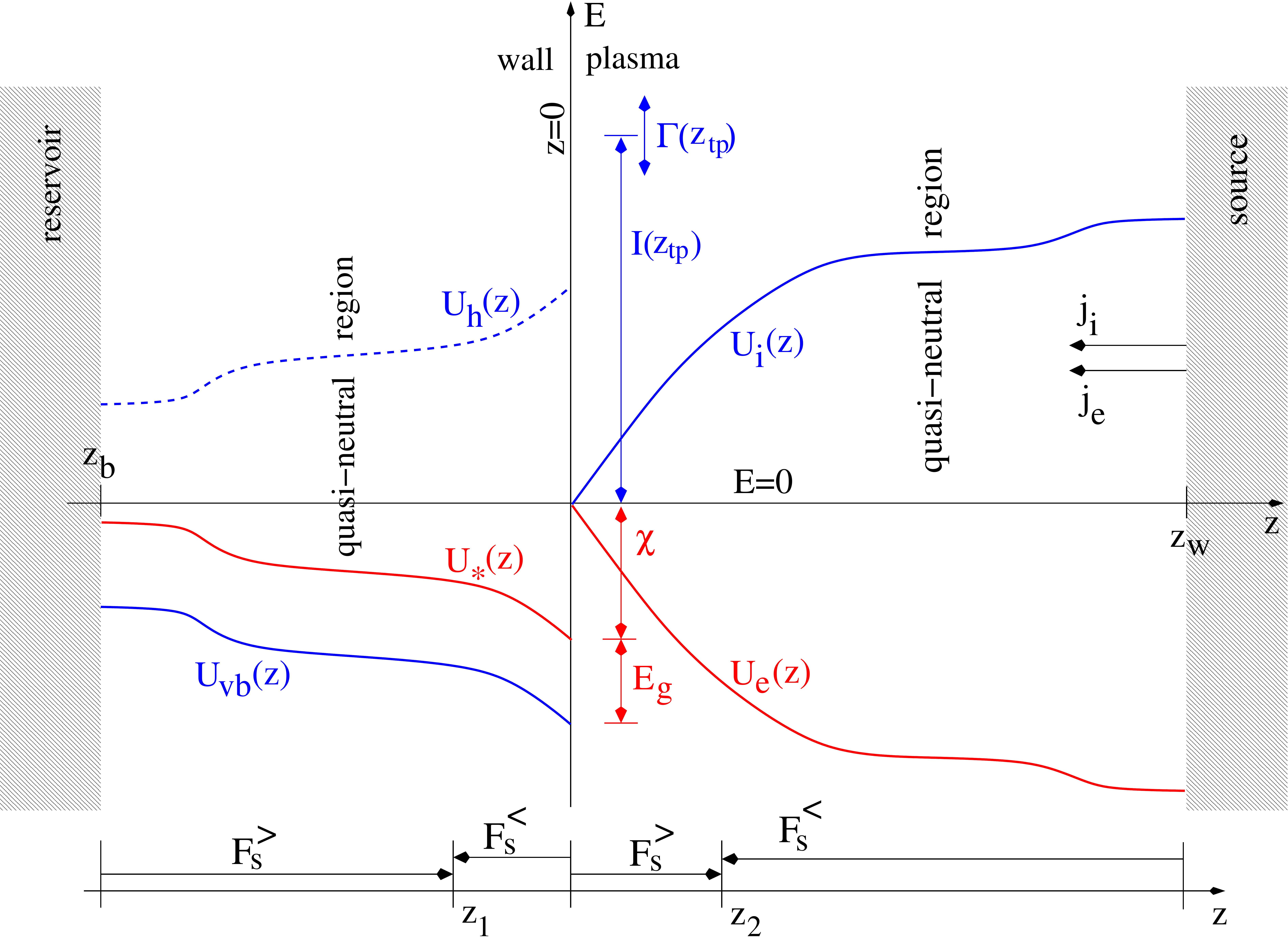}
\caption{(Color online) Illustration of the interface model on which our calculation is 
based anticipating an electric double layer with negative space charge inside the wall
and positive space charge in front of it. We consider a planar interface at $z=0$ separating
a dielectric wall from a plasma. The wall contains two bands, an initially fully occupied
valence band and an initially empty conduction band. Their edges $U_{\rm vb}$ and $U_*$ 
depend on $z$ because of the negative space charge accumulated from the plasma in response 
to the electron and ion fluxes originating from a plasma source at $z=z_w$. The blue dashed 
curve on the left indicates the $z-$dependent edge for valence band holes as explained in 
more detail in the main text. The plasma side shows the electric potential energy for electrons 
and ions arising from the positive space charge in front of the wall. In addition the ion's ionization 
level $I$ at the turning point $z_{\rm tp}$ and its broadening $\Gamma$ due to the hybridization 
with the surface are shown. Neutralization of ions impinging on the interface leads to the 
injection of valence band holes in an energy interval specified by $I$ and $\Gamma$. Far away 
from the interface at $z=z_b$ the wall is a reservoir for conduction band electrons and 
valence band holes. In the regions around the inflection points the system 
is quasi-neutral and field-free. The bottom of the figure shows how the species' 
distribution functions for left and right moving particles are determined at two particular 
locations $z=z_1$ and $z=z_2$.
}
\label{Model}
\end{figure}

\section{Kinetic theory}    
\label{Formalism}

\subsection{General approach}
\label{GeneralApproach}

Before we consider the simpler case of a collisionless, perfectly absorbing dielectric 
surface in contact with a collisionless plasma we describe in this subsection the kinetic
modeling of the electronic response of a dielectric plasma interface in broader terms.
A similar formulation could be worked out for a metallic plasma interface. As far as the 
modeling presented in this subsection is concerned, the main difference between metallic 
and dielectric plasma-facing solids is that for metals the neutralization of ions does 
not always lead to the injection of holes into a fully occupied valence band. Instead, 
if the ion's ionization energy is small enough holes are injected into the partially
filled conduction band, that is, into the same band into which also electrons are injected 
from the plasma. In this case, only the conduction band is involved, which moreover is 
partially filled. The hole representation we use in the following is then no longer 
advantageous and the theory on the solid side of the interface has to be formulated 
solely in terms of distribution functions for the conduction band electrons.

We consider a planar interface at $z=0$ separating 
a dielectric solid residing in the $z<0$ subspace from a plasma in the $z>0$ 
subspace. The theoretical treatment of the interface is based on the Poisson equation 
and two sets of Boltzmann transport equations operating respectively in the positive 
and negative half-spaces and describing in total four species: electrons and ions on 
the plasma side and conduction band electrons and valence band holes on the wall side. 
We use a species index $s=e,i,*,h$ to denote electrons, ions, conduction band electrons, 
and valence band holes. The interface is assumed to be homogeneous in the lateral 
directions $x$ and $y$ so that the spatial dependence of all quantities is reduced
to a dependence on $z$. For a quasi-stationary electric double layer at a homogeneous
interface the distribution functions of the various species depend thus only on the 
spatial coordinate $z$ and the three-dimensional wave vector $\vec{k}$.

Anticipating the quantum-mechanical derivation of the matching conditions for the 
distribution functions~\cite{DLP95,Falkovsky83} as well as a possibly iterative 
numerical treatment of the Boltzmann equations along the lines employed in the 
transport theory of semiconductor heterostructures~\cite{BW87,DP98} we replace 
the set of independent variables $(z,\vec{k})$ by $(z,E,\vec{K})$ where $E$ is 
the total energy and $\vec{K}$ the lateral momentum. The species' distribution 
functions are thus written as $F_s(z,E,\vec{K})$. 

If not stated otherwise, we give all equations in atomic units measuring energy 
in Rydbergs and length in Bohr radii. The zero of the energy scale is chosen to 
coincide with the electron affinity $\chi>0$ of the dielectric, that is, we set 
$E=\chi\equiv 0$.

The model on which our calculations rest is shown in Fig.~\ref{Model}. As it 
is drawn it already assumes an electric double layer with a negative and positive space
charge, respectively, inside the wall and inside the plasma. The double layer is driven 
by a source at $z=z_w>0$ releasing an electron flux $j_e$ and an ion flux $j_i$. We 
treat only the quasi-stationary case. The fluxes are thus equal to each other and 
exactly balanced by loss processes inside the wall. Far away from the interface, 
at $z=z_b<0$, the wall becomes a reservoir for conduction band electrons and valence 
band holes. 

On the wall side Fig.~\ref{Model} shows a valence and a conduction
band. Their edges are given by
\begin{eqnarray}
U_{\rm vb}(z) &= -U_c(z) - E_g - \chi~,  \qquad&  \label{Uvb}\\
U_{\rm *}(z) &= -U_c(z) -\chi &~  \label{Ucb}    
\end{eqnarray}
with $E_g$ the energy gap between the valence and the conduction band and 
$U_c(z)=eV_c(z)$ the electric potential energy given by the solution of the Poisson 
equation,
\begin{eqnarray}
\frac{d}{dz}\varepsilon(z)\frac{d}{dz}U_c(z)=8\pi \rho(z) ~, 
\label{Poisson}
\end{eqnarray}
where 
\begin{eqnarray}
\rho(z) = \rho_w(z)\theta(-z)-\rho_p(z)\theta(z) 
\label{RhoTotal1}
\end{eqnarray}
is the charge density and $\varepsilon(z)=\varepsilon\theta(-z)+\theta(z)$ is the dielectric 
constant, both split, with the help of the Heaviside function $\theta(z)$, into a wall and a 
plasma part. The connection between the solutions of the Poisson equation in the two 
half-spaces of the interface is given by the matching condition for the electric potential
energy, 
\begin{eqnarray}
U_c(0^-)=U_c(0^+)  ~~\mbox{\rm and}~~ \varepsilon\frac{d}{dz}U_c(0^-)=\frac{d}{dz}U_c(0^+)~.
\label{Umatch}
\end{eqnarray}
The way $\rho(z)$ is defined in (\ref{RhoTotal1}), the charge densities on the wall and 
the plasma side are given by 
\begin{eqnarray}
\rho_w(z) &=n_*(z)- n_D - n_h(z) + n_A ~,\qquad& \label{RhoWall}\\
\rho_p(z) &=n_i(z)-n_e(z)~, \label{RhoPlasma}& 
\end{eqnarray}
where $n_D$ and $n_A$ are, respectively, the concentration of donors and acceptors while 
$n_s(z)$ are the species's densities obtained by integrating the distribution 
functions over the independent variables $E$ and $\vec{K}$, 
\begin{eqnarray}
n_s(z)=\int\frac{dE d^2K}{(2\pi)^3} \frac{F_s(z,E,\vec{K})}{v_s(z,E,\vec{K})}~,
\label{Density}
\end{eqnarray}
where $v_s(z,E,\vec{K})$ is the absolute value of the $z-$component of the species' velocity. 
Since the distribution functions $F_s$ depend via the Boltzmann equation (see below) 
on $U_c$, the Poisson equation is in general a highly nonlinear integro-differential 
equation.

The physical meaning of the band edges is as follows: $U_*(z)$ gives the lowest energy 
a conduction band electron can have at $z<0$. Likewise $U_{\rm vb}(z)$ gives the highest 
energy a valence band electron can have at location $z<0$. At the plasma-wall interface 
valence band electrons per se are not directly relevant. What matters are the 
electrons in the valence band which have been extracted from it by neutralizing an ion 
impinging on the interface~\cite{MBF12}. It is thus natural to describe the valence band 
in terms of missing electrons, that is, in terms of holes~\cite{Hess88}. Instead 
of using electrons with a negative charge and a negative effective mass the hole picture
describes the valence band by quasi-particles with a positive charge and a positive 
effective mass. The energy a hole can have at location $z<0$ is always larger then 
\begin{eqnarray}
U_h(z)=-U_{\rm vb}=U_c(z)+Eg+\chi~,
\end{eqnarray}
indicated by the dashed blue line in Fig.~\ref{Model}. It is the edge for the motion of 
valence band holes.

On the plasma side the model contains ions and electrons. Their energies are given by
\begin{eqnarray}
U_{\rm i}(z) &= U_c(z) ~, \\
U_{\rm e}(z) &= -U_c(z) 
\end{eqnarray} 
with $U_c(z)$ the solution of the Poisson equation in the positive half-space. The energies
for ions and electrons at position $z>0$ are always larger than, respectively, $U_i(z)$ 
and $U_e(z)$. Figure~\ref{Model} also shows the ionization level $I$ of the ion and its 
broadening $\Gamma$, both taken at the turning point $z_{\rm tp}$ of the ion trajectory.
These two energies are needed in the hole injection model to be described later. The
ionization energy determines at what energy the hole is injected into the valence band 
and the broadening gives the energy range over which injection may occur. Notice, it is 
not the kinetic energy of the ion's center of mass motion which matters for hole injection 
but the ion's internal potential energy $I$.

At the bottom of Fig.~\ref{Model} we illustrate how distribution functions at particular 
locations, for instance, $z=z_1<0$ and $z=z_2>0$ can be determined if the distribution 
functions are known at the boundaries $z=z_b$ and $z=z_w$ and matching conditions 
are available connecting distribution functions across the interface at $z=0$. For 
each species $s$ we distinguish particles moving to the left from particles moving to the 
right. Hence, we write 
\begin{eqnarray}
F_s(z,E,\vec{K})=F_s^<(z,E,\vec{K}) + F_s^>(z,E,\vec{K})
\label{Split}
\end{eqnarray}
with $F_s^<$ and $F_s^>$ characterized, respectively, by $v_{z,s}=-v_s<0$ and $v_{z,s}=v_s>0$, 
where $v_s$ is the absolute value of the $z$-component of the species' velocity. Assuming for 
simplicity parabolic bands inside the dielectric, the velocities can be written as
\begin{eqnarray}
v_s(z,E,\vec{K}) = 2\bigg(\frac{m_e}{m_s}[E-U_s(z)] - (\frac{m_e}{m_s}\vec{K})^2 \bigg)^{1/2}~, 
\end{eqnarray}
where $m_e$ is the electron mass and $m_s$ is the species's (effective) mass.

Far away from the interface we assume the distribution functions to be local Maxwellians,
\begin{eqnarray}
\frac{F^{\rm LM}_s(z,E,\vec{K})}{n_s^{\rm LM}(z)}\!=\!\bigg(\frac{4\pi m_e}{k_BT_s m_s}\bigg)^{3/2}
                          \!\!\!\!\!\!\!\exp\bigg(\!\!\!-\frac{E-U_s(z)}{k_BT_s}\bigg) ,
\label{LM}
\end{eqnarray}
normalized to the density
\begin{eqnarray}
\int \frac{dE d^2K}{(2\pi)^3}\frac{F^{\rm LM}_s(z,E,\vec{K})}{v_s(z,E,\vec{K})}=n_s^{\rm LM}(z)~.
\end{eqnarray} 
The temperatures $T_s$ in (\ref{LM}) are input parameters whereas the values the densities 
$n_s^{\rm LM}(z)$ approach at the boundaries, the boundary densities at $z=z_b$ and $z=z_w$, 
are considered as variables. Hence, $n_{b*}=n^{\rm LM}_{*}(z_b)$, $n_{bh}=n^{\rm LM}_{h}(z_b)$,
$n_{se}=n^{\rm LM}_{e}(z_w)$, and $n_{si}=n^{\rm LM}_{i}(z_w)$ have to be determined in the 
course of the calculation. This adjustment of the densities is required to make the plasma
source consistent with the losses and the reservoir inside the wall mimicking the 
self-consistent response of the plasma to the wall and vice versa as it takes place in reality. 
On the plasma side this leads to the sheath modeling of Schwager and Birdsall~\cite{SB90} 
which has been also used in a number of particle-in-cell simulations~\cite{TLC04,GK12}. 

In order to determine $F_s^{\gtrless}$ in the way sketched at the bottom of Fig~\ref{Model}
we need matching conditions at $z=0$. Using the methods employed to match distribution 
functions across solid interfaces, originally developed by Falkovsky~\cite{Falkovsky83} for 
charge transport inside metallic surfaces, but subsequently applied also to solid 
heterostructures~\cite{DLP95,Schroeder92}, the distribution functions for electrons in the 
plasma and electrons in the conduction band of the wall are for $E>0$ at $z=0$ connected by
\begin{eqnarray}
F_e^>(0,E,\vec{K}) &= R(E,\vec{K}) F_e^<(0,E,\vec{K}) & \nonumber\\
                   &+ [1-R(E,\vec{K})]F_*^>(0,E,\vec{K})~, \qquad& \label{Ematch1}\\
F_*^<(0,E,\vec{K}) &= R(E,\vec{K}) F_*^>(0,E,\vec{K}) & \nonumber\\
                   &+ [1-R(E,\vec{K})]F_e^<(0,E,\vec{K}) \qquad&
\label{Ematch2}
\end{eqnarray}
with 
\begin{eqnarray}
R(E,\vec{K})=\bigg(\frac{v_e(0,E,\vec{K})-v_*(0,E,\vec{K})}{v_e(0,E,\vec{K})+v_*(0,E,\vec{K})}\bigg)^2
\end{eqnarray}
the quantum-mechanical reflection probability for a three-dimensional potential step
of height $\chi$. For energies just above the potential step $R$ is close to unity while 
for large $E$ it vanishes. 

It should be noted that $R$ describes only the quantum-mechanical specular reflection on the 
potential step. Inelastic processes inside the wall which may bring the electron back to the 
interface, and hence also possibly back to the plasma, if it successfully passes the potential 
step in the reverse direction, are not included in $R$. Hence, in (\ref{Ematch1}) and 
(\ref{Ematch2}) the reflection coefficient $R$ cannot be replaced by $1-S$ with $S$ the electron
sticking coefficient obtained, for instance, by the method of invariant embedding~\cite{BF15}
because $S$ accounts for inelastic processes which in the present modeling of the interface's
particle kinetics have to be incorporated into the collision integrals of the Boltzmann equations. 
The same holds for secondary electron emission due to impacting electrons with energies above 
the band gap of the wall. It also arises from inelastic processes which have to be accounted for 
by the collision integrals of the Boltzmann equations. Secondary electron emission due to Auger 
neutralization of ions~\cite{PP99} or other heavy particle collisions with the surface, on other 
hand, could be incorporated into the matching conditions by augmenting the right hand side of 
(\ref{Ematch1}) by a source function $S_e^>(E,\vec{K})$ describing the probability of the surface 
to emit an electron with total energy $E$ and lateral momentum $\vec{K}$ because the kinetic equations 
we set up do not track the occupancies of the internal electronic states of the ion (or other 
heavy projectiles). The particular form of this function depends on the elementary surface process. 
It is best worked out semi-empirically along the lines we use now for the resonant neutralization of 
ions at a surface.

The matching of the ion and hole distribution functions differs from the matching of electron 
distribution functions because ions cannot enter the wall and most importantly the center of 
mass motions of the ions and the valence band holes are not coupled via energy and momentum 
conservation laws. In our model ions are resonantly neutralized at the interface whereby 
they inject holes into the wall's valence band. Hence, the ion and hole distribution 
functions are for $E>E_g+\chi$ at $z=0$ connected by 
\begin{eqnarray}
F_h^<(0,E,\vec{K}) &= F_h^>(0,E,\vec{K}) + \alpha S_h^<(E,\vec{K})~, \label{Imatch1}\\
F_i^>(0,E,\vec{K}) &= (1-\alpha)F_i^<(0,E,\vec{K}) \label{Imatch2}
\end{eqnarray}
with $S_h^>$ a source function to be constructed by other means, $\alpha$ the 
probability for wall neutralization, which we assume here to be independent of the energy 
of the ion's center of mass motion. If the ions are neutralized with probability $\alpha$ 
less than unity some ions have the chance to come back to the plasma, having thus velocity
$v_i$ in negative $z$-direction and contributing thereby to $F_i^<$. Only for $\alpha=1$, 
that is, for perfect neutralization, no ions are coming back to the plasma, as it is  
often assumed in the modeling of the plasma sheath. 

A model for hole injection is required to complete the description of the matching condition 
(\ref{Imatch1}). The source function $S_h^<(E,\vec{K})$ is the probability to inject an 
hole with total energy $E$ and lateral momentum $\vec{K}$ into the valence band. It is 
important to realize that in~(\ref{Imatch1}) $E$ and $\vec{K}$ are not the total energy 
and lateral momentum of the impinging ion responsible for the injection. To connect 
the source function $S_h^<(E,\vec{K})$ with the ion distribution function at the 
interface, $F_i^<(0,z,E)$, we recall that the total flux of injected holes has to be 
identical to the total flux of impinging ions multiplied by $\alpha$. Hence, with 
\begin{eqnarray}
j_s(z)=\int \frac{dE d^2K}{(2\pi)^3} [F_s^>(z,E,\vec{K})-F_s^<(z,E,\vec{K})]~,
\label{DefFlux}
\end{eqnarray}
we obtain by using (\ref{Imatch2}) the condition
\begin{eqnarray}
\int \frac{dE d^2K}{(2\pi)^3} F_i^<(0,E,\vec{K})=\int\frac{dE d^2K}{(2\pi)^3} S_h^<(E,\vec{K})~,
\label{Source}
\end{eqnarray}
where the integration variables on the left (right) belong to the center of mass 
motion of the ion (hole). To proceed one either determines $S_h^<(E,\vec{K})$ from 
a microscopic model for ion neutralization at a surface or one makes plausible 
assumptions about the overall behavior of this function. In the next section, where 
we discuss the simple collisionless, perfectly absorbing interface, we assume, for
instance, holes to be injected with uniform probability over the relevant energy and 
momentum range. 

The matching conditions (\ref{Ematch1}), (\ref{Ematch2}), (\ref{Imatch1}), and (\ref{Imatch2}) 
are essential for our approach. A comment about their impact on the distribution functions is 
thus in order. The distribution functions react freely to the matching conditions, that is, 
the values they assume at $z=0$ are determined self-consistently by the interplay of the 
plasma with the solid. Close to the interface the distribution 
functions deviate from the distribution functions far away from it in precisely such a way as 
it is dictated by the matching conditions. Since the matching occurs predominately in the tails
of the distribution functions, that is, at high energies, where charge carriers cross the
interface, collisions have enough phase space to efficiently heal the perturbation due to 
the interface making a merging of the solid and plasma distribution functions possible. 

Let us now turn to the (quasi-stationary) Boltzmann equations to be satisfied by the distribution 
functions of the four species. They can be cast 
into the form
\begin{equation}
\left[\pm v_s(z,E,\vec{K})\frac{\partial}{\partial z}+\gamma_s[F_{s^\prime}^{\gtrless}]\right]
F_s^{\gtrless}(z,E,\vec{K})=\Phi_s[F_{s^\prime}^{\gtrless}]~,
\label{BTE}
\end{equation}
where $z$ is either positive or negative, depending on the species, and 
$\gamma_s[F_{s^\prime}^{\gtrless}]$ and $\Phi_s[F_{s^\prime}^{\gtrless}]$ are collision 
integrals which also depend on the species. For instance, ions 
may suffer charge-exchange collisions whereas electrons may be collisionless on the 
plasma side but subject to intra- and interband Coulomb and phonon collisions on the 
wall side. Similarly holes may also suffer intra- and interband Coulomb and phonon collisions.
In addition, electron-hole recombination may take place inside the wall, for instance, via Auger
or radiative processes. All of it has to be included in the collision integrals $\gamma_s$ and $\Phi_s$. 
Since in the following we numerically treat only the collisionless interface we do not give 
explicit formulae for $\gamma_s$ and $\Phi_s$ but they can be worked out in all cases using 
the techniques of semiclassical kinetic theory~\cite{SJ89}.

Once the collision integrals are specified the set of equations for the theoretical description 
of the electric double layer is complete. It contains the modifications of the band structure 
inside the dielectric wall as well as the modification of the electric potential in front of 
the wall due to the permanent influx of electrons and ions. The fluxes are released from a 
plasma source and annihilated inside the wall. How the fluxes are annihilated, radiatively 
or non-radiatively, is beyond the present model. Solving the Poisson equation(\ref{Poisson})
together with the Boltzmann equations (\ref{BTE}) for the four species subject to the matching
conditions (\ref{Umatch}), (\ref{Ematch1}), (\ref{Ematch2}), (\ref{Imatch1}), and (\ref{Imatch2})
gives the species' distribution functions and eventually the density and potential profiles of 
the double layer across the interface. The source and the reservoir can be made self-consistent 
by enforcing additional conditions to fix the boundary densities $n_{b*}$, $n_{bh}$, $n_{se}$, 
and $n_{si}$,  which depend however on what kind of collisions are included. This in turn 
determines the way the system of Boltzmann-Poisson equations is solved numerically. In 
the next subsection we treat the simplest case--a collisionless interface--in full detail. It 
thus becomes apparent what additional conditions are needed and how the Boltzmann-Poisson 
equations are actually solved.

\subsection{Collisionless, perfectly absorbing surface}

So far the description of our approach was quite general. We now specialize the treatment
to the simplest possible case: a collisionless plasma in contact with a perfectly absorbing
collisionless surface. The reason is a practical one. In this particular case the 
Boltzmann equations become first order ordinary differential equations from which the
two-dimensional lateral momentum $\vec{K}$ can be eliminated. Since $\vec{K}$ 
also drops out from the perfect absorber matching conditions, one no longer deals with
distribution functions depending on four independent variables $(z,E,\vec{K})$ but only 
on two $(z,E)$. Without collision integrals the Boltzmann equations can be furthermore 
easily integrated yielding, in addition, charge densities which do not explicitly depend 
on $z$ but only on $U_c(z)$. The Poisson equation can thus be integrated once analytically 
and the remaining numerical task is rather modest. In Sect.~\ref{Results} we will discuss
how realistic the collisionless theory is.

The lateral momentum can be eliminated as follows. First, the collisionless Boltzmann 
equations are integrated over the lateral momentum $\vec{K}$ before switching from 
$(z,\vec{k})$ to $(z,E,\vec{K})$ as independent variables. They then become equations for 
\begin{eqnarray}
F_s(z,k)=\int\frac{d^2K}{(2\pi)^2} F_s(z,\vec{K},k) ~.
\end{eqnarray}
Switching then from $(z,k)$ to $(z,E)$, with $E$ now the total energy without the kinetic
energy in the lateral directions, leads to collisionless Boltzmann equations of the form 
\begin{equation}
\pm v_s(z,E)\frac{\partial}{\partial z}F_s^{\gtrless}(z,E)=0
\label{BTECollLess}
\end{equation}
from which the distribution functions $F_s^\gtrless(z,E)$ can be determined. Since $E$
no longer contains the kinetic energy in the lateral directions, the velocities in 
(\ref{BTECollLess}) are given by 
\begin{eqnarray}
v_s(z,E) = 2\bigg(\frac{m_e}{m_s}[E-U_s(z)]\bigg)^{1/2}
\end{eqnarray}
with $U_s(z)$ still defined as before.

Next, setting $R=0$ for a perfectly absorbing surface, the matching conditions for the 
electron distribution functions, (\ref{Ematch1}) and (\ref{Ematch2}), can be transformed
back to $\vec{k}$-space and integrated there over $\vec{K}$ leading to conditions for 
$F_{*,e}^\gtrless(z,k)$ . Changing now the 
independent variables $(z,k)$ to $(z,E)$ as in the Boltzmann equations leads to conditions 
for $F_{*,e}^\gtrless(z,E)$ with $E$ again the total energy without the kinetic energy in 
lateral directions. Obviously, this procedure is only applicable if $R$ is assumed to 
be independent of $\vec{k}$. In general this is not the case. Hence, to reduce in the 
general matching conditions the number of independent variables requires 
additional assumptions about the energy and momentum dependence of $R$. Only for 
the perfect absorber model, where $R=0$ from the start, no further assumptions are necessary.
For $\alpha=1$, as again postulated by the perfect absorber model, the matching conditions for 
ions can be handled similarly. Hence, the lateral momentum can be also eliminated from them.

\begin{figure}[t]
\includegraphics[width=0.99\linewidth]{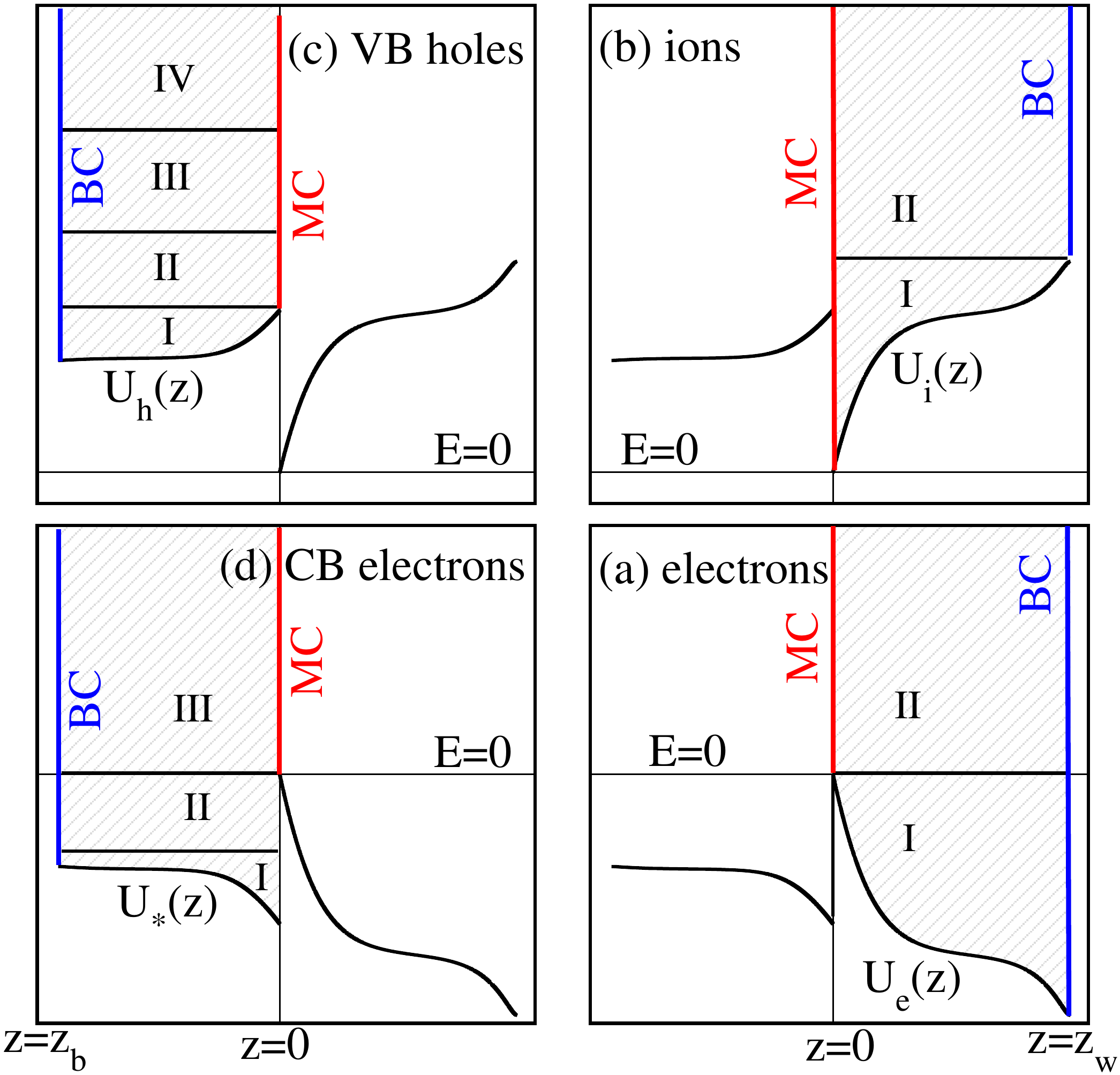}
\caption{(Color online) The panels show the domains of quasi-free motion of 
electrons (a), ions (b), valence band holes (c), and conduction band electrons 
(d) between the boundaries at $z=z_w$ or $z=z_b$, depending on the type of 
particle considered, and the interface at $z=0$. The boundary conditions at 
$z=z_w$ and $z=z_b$ and the matching conditions at $z=0$ are indicated by 
blue and red vertical lines. For energies where no matching conditions 
apply specular reflection occurs. As an example let us consider panel (a)
for the electrons on the plasma side. Due to the absence of collisions the 
function $F_{e}^<(z)$ is given for $E>U_e$ by $F_{e}^{\rm LM}(z_w)$ 
while the function $F_{e}^>(z)=F_e^<(z)=F_{e}^{\rm LM}(z_w)$ for $E<0$ 
(region I) but identical to the function $F_e^>(0,E)$ given by the matching 
conditions at $z=0$ for $E>0$ (region II). A similar analysis applies to the 
other species as well. One only has to remember that regions not directly 
connected to the boundary conditions cannot contain any particles because of 
the lack of collisions. For instance, region I in panel (d) cannot be populated 
by conduction band electrons. Likewise region I in panel (b) is void of ions. 
Taking these considerations into account the solutions of (\ref{BTECollLess}) 
can be constructed.}
\label{TA}
\end{figure}

Altogether, in terms of the functions $F_s^\gtrless(0,E)$ the matching conditions for the 
collisionless, perfectly absorbing interface become 
\begin{eqnarray}
F^>_e(0,E) &=F^>_*(0,E)=0 ~,\label{match1}\\
F^<_*(0,E) &=F_e^<(0,E) ~,\label{match2}\\
F^<_h(0,E) &= F^>_h(0,E) + S_h^<(E)~,\label{match3}\\
F^>_i(0,E) &= 0 ~.\label{match4}
\end{eqnarray}
The first equation indicates that for electrons not to come back to the plasma, as 
postulated by the perfect absorber model, we have to assume not only $R=0$ but in 
addition $F^>_*(0,E)=0$ for $E>0$. This thermalization condition is necessary because
$R$ is the same in Eqs. (\ref{Ematch1}) and (\ref{Ematch2}). Quantum mechanical 
reflection by a potential step does not depend on the direction the step is crossed. 
Conditions (\ref{match3}) and (\ref{match4}) describe a surface perfectly annihilating
ions by injecting holes into the valence band. Assuming the source function $S_h^<(E)$ 
to be a uniform probability for hole injection, it is given by 
\begin{eqnarray}
S_h^<(E)=\bar{S}[\theta(E-I+\Gamma/2)-\theta(E-I-\Gamma/2)]
\end{eqnarray}
with the normalization 
\begin{eqnarray}
\bar{S}=\frac{1}{\Gamma}\int_{U_w}^0 dE F_i^<(0,E)
\label{Sbar}
\end{eqnarray}
to ensure, at the interface, the equality of ion and hole flux. In the normalization 
condition for $S_h^<(E)$ we already anticipated for $F_i^<(0,E)$ the range of 
integration discussed in Fig. \ref{TA} and set $U_w=U_c(z_w)$.  

Equations (\ref{BTECollLess}) can be solved by a trajectory analysis
taking matching and boundary conditions into account as illustrated in Fig.~\ref{TA} .
In the collisionless model the boundary conditions at $z=z_b$ and $z=z_w$ are given 
by the Maxwellians 
\begin{eqnarray}
\frac{F_s^{\rm LM}(z,E)}{n^{\rm LM}_s(z)}\!=\!\bigg(\frac{4\pi m_e}{k_BT_s m_s}\bigg)^{1/2}
                 \!\!\!\!\!\!\exp\bigg(\!\!-\frac{E-U_s(z)}{k_BT_s} \bigg) ~ 
\end{eqnarray}
as follows: At $z=z_b$ we enforce $F_{*,h}^>(z_b,E)=F_{*,h}^{\rm LM}(z_b,E)$ for 
$E>U_{*,h}(z_b)$ whereas at $z=z_w$ we require $F_{e,i}^<(z_w,E)=F_{e,i}^{\rm LM}(z_w,E)$ 
for $E>U_{e,i}(z_w)$. This is illustrated in Fig.~\ref{TA} by the vertical blue lines. 
The densities at $z=z_b$ and $z=z_w$, the boundary densities, given by 
$n_{b*}=n^{\rm LM}_{*}(z_b)$, $n_{bh}=n^{\rm LM}_{h}(z_b)$,
$n_{se}=n^{\rm LM}_{e}(z_w)$, and $n_{si}=n^{\rm LM}_{i}(z_w)$ 
are as pointed out in the previous section variables to be determined by the kinetic 
model. In addition to the boundary conditions the solution of (\ref{BTECollLess}) 
requires the matching conditions at $z=0$ symbolized in Fig.~\ref{TA} by the vertical 
red lines. Since the trajectory analysis is standard~\cite{HZ66} it is not explicitly
given here. The principle of the calculation is described in the caption
of Fig.~\ref{TA}.

It turns out that the densities obtained from the solutions of (\ref{BTECollLess}) 
depend only on $U_c(z)$ and not on $z$ explicitly. This greatly simplifies the further 
processing of the densities. In particular, it allows us to obtain the first integral 
of the Poisson equation analytically. What remains to be done numerically is only the 
calculation of the second integral of the Poisson equation and the solution of four 
nonlinear algebraic equations. This is of course much easier than a numerical solution 
of a set of collisional Boltzmann-Poisson equations. In our view this justifies working
out the collisionless, perfectly absorbing interface. Despite its idealistic nature 
the numerical solution may lead to insights useful for attacking the full problem.

Let us first consider the fluxes obtained from the solutions of (\ref{BTECollLess}).
Using the definition (\ref{DefFlux}) with the obvious modifications arising from the 
fact that we eliminated the lateral momentum $\vec{K}$ we find
\begin{eqnarray}
j_* &= j_e = n_{se} \bigg(\frac{k_BT_e}{\pi}\bigg)^{1/2}\exp\bigg(-\frac{U_w}{k_BT_e}\bigg) ~,\\
j_h &= j_i = n_{si} \bigg(\frac{m_e}{\pi}\frac{k_BT_i}{m_i}\bigg)^{1/2} 
\end{eqnarray}
for the fluxes on the plasma and the wall side of the double layer. Hence, the matching 
conditions and the hole injection model preserve by construction the fluxes 
across the interface. At quasi-stationarity, the electron and ion fluxes satisfy moreover 
the flux balance condition,
\begin{eqnarray}
j_i=j_e ~,
\label{FluxBalance}
\end{eqnarray}
leading to a first condition involving densities at the system's boundaries 
\begin{eqnarray}
\frac{n_{si}}{n_{se}}= \bigg(\frac{k_BT_e}{k_BT_i}\frac{m_i}{m_e}\bigg)^{1/2}\exp\bigg(-\frac{U_w}{k_BT_e}\bigg) ~.
\label{Alpha}
\end{eqnarray}

The densities $n_s(z)=n_s(U_c(z))$ are obtained from (\ref{Density}) using (\ref{Split}), again with 
the obvious modifications arising from the elimination of $\vec{K}$. With the expressions for $n_s(U_c)$ 
given in \ref{Functions} the source for the Poisson equation on the plasma side becomes  
\begin{eqnarray}
\rho_p(z)=\rho_p(U_c)=n_i(U_c)-n_e(U_c)
\label{RhoPlasma1}
\end{eqnarray} 
while on the wall side the source is given by 
\begin{eqnarray}
\rho_w(z)=\rho_w(U_c)=\rho_w^t(U_c)+\rho_w^j(U_c) 
\label{RhoWall1}
\end{eqnarray}
with
\begin{eqnarray}
\rho_w^t(U_c) = n_*^t(U_c) - n_D - n_h^t(U_c) + n_A ~, \label{RhoWallt}\\
\rho_w^j(U_c) = n_*^j(U_c) - n_h^j(U_c) \label{RhoWallj} ~.
\end{eqnarray}

In the formulas given above, the densities $n_{*,h}^t(U_c)$ describe conduction band electrons 
and valence band holes which are thermalized/trapped within the wall while the densities
$n_{*,h}^j(U_c)$ originating from (\ref{match1}) and (\ref{match2}) describe carriers coming from the 
continuing influx of electrons and ions to the interface after quasi-stationarity has been 
reached. This influx does not stop after the double layer is fully developed. Quasi-stationarity 
makes the electron and ion fluxes coincide, rather than vanish. In the expressions for 
$n_{*,h}^j$ we employed already (\ref{FluxBalance}) to replace the fluxes $j_*$ and $j_h$ by the 
ion density $n_{si}$.

Since the Poisson equation's sources $\rho_{p,w}$ depend only on $U_c$ it can be integrated once in 
each half-space. Let us first consider the plasma side. Multiplying (\ref{Poisson}) by $U_c^\prime$, 
where the prime indicates here and in the formulae to follow a derivative with respect
to $z$, and using $U_c(0^+)=0$, together with $U_c^\prime(z_w)=0$, which forces the double 
layer to be field-free at $z=z_w$, yields
\begin{eqnarray}
U_c^\prime(z)={\cal F}(U_c) ~,
\label{PoissonPlasma}
\end{eqnarray}
where 
\begin{eqnarray}
{\cal F}(U_c)= \bigg( 16\pi \int_{U_c}^{U_w} d\bar{U}_c\rho_p(\bar{U}_c)  \bigg)^{1/2}
\label{SagdeevF}
\end{eqnarray}
with $U_w=U_c(z_w)$. In~\ref{Functions} the result of this integration is given. 
The potential profile on the plasma side can thus be obtained  
by integrating (\ref{PoissonPlasma}) from $z=0$ to $z>0$. The result is
\begin{eqnarray}
\int_0^{U_c} \frac{d\bar{U}_c}{{\cal F}(\bar{U}_c)}=z  
\label{UcPlasma}
\end{eqnarray}
with $0 \le z \le z_w$.

In our model we assume $\rho_w^t$ to act inside the wall as the only source for the 
electric potential energy $U_c$. It is this part of the wall's charge density which 
balances the positive charge density $\rho_p$ on the plasma side of the interface. 
The density $\rho_w^j$ acts in our model not as a source. Instead it will be made 
to vanish (physically due to electron-hole recombination inside the wall, see below). 
Thereby it yields an additional condition which in conjunction with the other 
conditions to be satisfied at the interface enables us to calculate also the continuing 
influx of electrons and ions. Using thus inside the wall $\rho_w^t$ as the only source for 
$U_c$ and a procedure similar to the one employed to derive (\ref{PoissonPlasma}) leads 
us to 
\begin{eqnarray}
U_c^\prime(z)={\cal G}^t(U_c) ~,
\label{PoissonWall}
\end{eqnarray}
where 
\begin{eqnarray}
{\cal G}^t(U_c)= \bigg( \frac{16\pi}{\varepsilon} \int_{U_b}^{U_c} d\bar{U}_c\rho^t_w(\bar{U}_c) \bigg)^{1/2}
\label{SagdeevG}
\end{eqnarray}
and $U_b=U_c(z_b)$. The function ${\cal G}^t(U_c)$ is given in~\ref{Functions}.
In deriving these expressions we used $U_c(0^-)=0$ which guarantees continuity of 
$U_c$ at $z=0$, as required by the first condition in (\ref{Umatch}). In addition we forced
the double layer to be field-free at $z=z_b$ by setting $U_c^\prime(z_b)=0$. The potential 
profile on the wall side is given by integrating (\ref{PoissonWall}) from $z<0$ to 
$z=0$ resulting in 
\begin{eqnarray}
\int_{U_c}^0 \frac{d\bar{U}_c}{{\cal G}^t(\bar{U}_c)}=-z
\label{UcWall}
\end{eqnarray}
with $z_b \le z \le 0$.

In order to incorporate into the formalism the role we want the densities $n_{*,h}^j$ 
to play, we now take a closer look at them. As already mentioned they arise from 
the continuing influx of electrons and ions after the quasi-stationarity of the double 
layer has been reached. Hence, for a quasi-stationary double layer $n_{*,h}^j$ cannot 
act as a source for an electric field as we already anticipated in (\ref{PoissonWall}). 
It is thus reasonable to assume that they recombine nearby the interface, perhaps in a 
spatial zone stretching from $z=0$ to $z=z_1<0$, where $z_1$ is a distance from the 
interface not yet specified. Below we will assume $z_1$ to coincide with the inflection 
point of $U_c(z)$ inside the wall which we need to match the double layer to a quasi-neutral,
field-free region, as it happens in reality. On the plasma side an inflection point 
has to be implemented for the same reason. There it is required to match the sheath to a 
quasi-neutral, field-free plasma~\cite{SB90}. 

In view of what we just said, we hence postulate the recombination condition,  
\begin{eqnarray}
\int_{z_1}^0 dz \rho_w^j(z) = \int_{U_1}^0 dU_c\frac{\rho_w^j(U_c)}{{\cal G}^t(U_c)} = 0 ~  
\label{ReComb}
\end{eqnarray}
with $U_1=U_c(z_1)$ the electric potential energy at the inflection point $z=z_1$. In
the second equality we used (\ref{PoissonWall}) to replace the $z-$integration by an 
integration over $U_c$. The last condition we finally have to enforce is the jump condition 
for the derivative of the electric potential energy at $z=0$ as stated in (\ref{Umatch}). 
Using~(\ref{PoissonPlasma}) and~(\ref{PoissonWall}) it becomes 
\begin{eqnarray}
\varepsilon {\cal G}^t(0)={\cal F}(0) ~
\label{Kappa}
\end{eqnarray}
with ${\cal F}(U_c)$ and ${\cal G}^t(U_c)$ defined in~(\ref{SagdeevF}) and~(\ref{SagdeevG}).

Now, we have all the ingredients together to formulate a self-consistent model for the 
electric double layer at a collisionless, perfectly absorbing plasma-wall interface. 

Towards that end let us first discuss the necessity of implementing inflection 
points into the potential profile $U_c(z)$. It can be most clearly seen by setting 
hypothetically $n_{si}=n_{se}$ and $n_{b*}=n_{bh}$, that is, by making the source
and the reservoir charge-neutral. Such a choice would however not lead to 
$\rho_p(z_w)=0$ and $\rho_w(z_b)=0$ as one would perhaps naively expect. Hence, 
by initially assuming at $z=z_b$ and $z=z_w$ distribution functions $F_s^{\rm LM}$ 
confronts us at the end with charge non-neutral 
boundaries if at the same time we force the net charge of the boundary densities to be 
zero. Hence, $F_s^{\rm LM}$ cannot describe the quasi-neutral regions into which 
the double layer should be embedded. 

The reason is of course that some particles are, depending on their energy and/or 
type, either absorbed or emitted by the interface. Hence, they are lost from or 
gained by a half-space of the double layer preventing thereby $F_s^\gtrless$ to 
re-establish $F_s^{\rm LM}$ at the boundaries. In reality the distribution functions 
react self-consistently to the presence of the interface making thereby the double 
layer also charge-neutral far away from the interface. By postulating the form of the 
distribution functions at $z=z_b$ and $z=z_w$ we destroyed this mechanism. Alternatively 
we could have enforced charge-neutrality at the boundaries. But then we could not 
have known the distribution functions making the solution of (\ref{BTECollLess}) 
much more complicated.

To mimic the self-consistent reaction of the distribution functions far away 
from the interface we follow Schwager and Birdsall~\cite{SB90} and consider the 
boundary densities, $n_{b*}$, $n_{bh}$, $n_{se}$, and $n_{si}$, appearing in $F_s^{\rm LM}$ 
at $z=z_b$ and $z=z_w$, respectively, as variables to be determined from the calculation. 
This requires to incorporate two inflection points into the potential profile, one at 
$z=z_1<0$ inside the wall and one at $z=z_p>0$ inside the plasma. The conditions for the 
inflection points are charge-neutrality 
\begin{eqnarray}
\rho^t_w(U_1)=0 ~, \label{NeutralW1a}\\
\rho_p(U_p)=0 \label{NeutralP1}
\end{eqnarray}
and the vanishing of the electric field 
\begin{eqnarray}
{\cal G}^t(U_1)=0 ~, \label{FieldFreeW} \\
{\cal F}(U_p)=0 \label{FieldFreeP} ~
\end{eqnarray}
with $U_1=U_c(z_1)$ and $U_p=U_c(z_p)$. Notice, on the 
wall side only ${\cal G}^t$ appears. 

The charge-neutrality conditions (\ref{NeutralW1a}) and (\ref{NeutralP1}) have to be
specified further. On the plasma side it translates simply into 
\begin{eqnarray}
n_e(U_p)=n_i(U_p)~.
\label{NeutralP2}
\end{eqnarray}
On the wall side, however, charge-neutrality is more involved since in addition to 
(\ref{NeutralW1a}) we also have to satisfy 
\begin{eqnarray}
n^t_*(U_1) n^t_h(U_1)=n_{\rm i}^2
\label{NeutralW1b}
\end{eqnarray}
with
\begin{eqnarray} 
n_{\rm i} = \frac{1}{4}\bigg(\frac{k_BT_*}{\pi}\bigg)^{3/2}
      \bigg(\frac{m_*m_h}{m_e^2}\bigg)^{3/4}\exp\bigg(-\frac{E_g}{2 k_BT_*}\bigg)
 \label{nint}
\end{eqnarray} 
the intrinsic charge density~\cite{Li06} of the wall at temperature $T_*=T_h$. 
Using the formulae for the densities given in \ref{Functions} and solving 
Eqs. (\ref{NeutralW1a}) and (\ref{NeutralW1b}) simultaneously, assuming either 
$n_A=n_D=0$ (intrinsic wall), $n_A=0$ (n-doped wall), or $n_D=0$ (p-doped wall) 
we obtain two conditions for the boundary densities $n_{b*}$ and $n_{bh}$:
\begin{eqnarray}
\frac{n_{b*}}{n_{\rm ref}} = N^2_*\frac{\exp\bigg(\frac{U_b-U_1}{k_BT_*}\bigg)}
{\Phi\bigg(\sqrt{\frac{U_1+\chi}{k_BT_*}}\bigg)-\Phi\bigg(\sqrt{\frac{U_1-U_b}{k_BT_*}}\bigg)} 
\label{Beta}
\end{eqnarray}
and 
\begin{eqnarray}
\frac{n_{bh}}{n_{\rm ref}} = N^2_h\exp\bigg(\frac{U_1-U_b}{k_BT_h}\bigg) ~,
\label{Gamma}
\end{eqnarray}
where $\Phi(\sqrt{y})$ is the error function, see \ref{Functions}. For an 
intrinsic wall $n_{\rm ref}=n_{\rm int}$ and $N_*=N_h=1$, for a p-type wall
$n_{\rm ref}=n_p=[n_A+\sqrt{n_A^2+4n_i^2}]/2$ leading to $N_*=2x/(1+\sqrt{1+4x^2})$ 
and $N_h=1$ with $x=n_{\rm int}/n_A$, while for a n-type wall 
$n_{\rm ref}=n_n=[n_D+\sqrt{n_D^2+4n_i^2}]/2$ yielding $N_*=1$ and $N_h=2x/(1+\sqrt{1+4x^2})$ 
with $x=n_{\rm int}/n_D$. From the jump condition (\ref{Kappa}) we finally obtain an
equation relating $n_{si}/n_{\rm ref}$ to the density ratios $n_{b*}/n_{\rm ref}$ 
and $n_{bh}/n_{\rm ref}$ making the approach self-consistent.  

The description of our approach is now complete. The modeling we propose for a 
quasi-stationary electric double layer at a collisionless, perfectly absorbing plasma-wall 
interface contains eight parameters: Four energies $U_b$, $U_1$, $U_p$, and $U_w$ and four 
densities $n_{b*}$, $n_{bh}$, $n_{se}$, and $n_{si}$. Eight equations are available to 
determine them: The three conditions (\ref{Alpha}), (\ref{Beta}), and (\ref{Gamma})
for the boundary densities, the quasi-neutrality condition on the plasma side 
(\ref{NeutralP2}), the two conditions forcing the double layer
to be field-free around $z=z_1$ and $z=z_p$, (\ref{FieldFreeW}), 
(\ref{FieldFreeP}), the recombination condition (\ref{ReComb}) and the 
jump condition (\ref{Kappa}) guaranteeing at the end that the double layer 
is globally charge neutral between its physically relevant boundaries $z_1$ and 
$z_p$. It should be noticed that the wall provides an absolute scale via the 
reference density $n_{\rm ref}$ and the band structure parameters $E_g$ and 
$\chi$, as does the ionization energy $I$ and its broadening $\Gamma$. The
approach produces thus absolute numbers. 


\section{Results}
\label{Results}

In this section we use parameters applicable to \Si\ and \SiOTwo\ surfaces in 
contact with a two-temperature hydrogen plasma to obtain numerical results for 
the electric double layer forming at a collisionless, perfectly absorbing plasma-wall 
interface. Before discussing the results we give some details about the numerical 
treatment of the equations derived in the previous section.

\begin{table}[t]
  \begin{tabular}{|l||llllll|}
    \hline
    Wall      & $\frac{m_e}{m_h}$        & $k_BT_*$      & $k_BT_h$ & $\chi$         & $E_g$      & $\varepsilon$ \\
              &                          & (eV)          & (eV)     & (eV)           & (eV)       &               \\\hline
    \Si       & 1.0                      & 0.2           & 0.2      & 4              & 1.0        & 12            \\
    \SiOTwo   & 1.0                      & 0.2           & 0.2      & 1.0            & 9.0        & 4             \\\hline
              &                          &               &          &                &            &               \\
    Plasma    & $\frac{m_e}{m_i}$        & $k_BT_e$      & $k_BT_i$ & $I$            & $\Gamma$   &               \\
              & ($10^{-4}$)              & (eV)          & (eV)     & (eV)           & (eV)       &               \\\hline
    ${\rm H}^+$-e        & 5.4                      & 2.0           &  0.2     & 13.6           & 2.0        &               \\\hline
  \end{tabular}
\caption{Material parameters for the wall and the plasma used in the numerical calculations.
In order not to overload the model we neglect the image shift of the ions' ionization level.
It would depend on yet another parameter, the position of the image plane. The mass of the
conduction band electrons $m_*$ is not included in the table because it is varied between
physically reasonable bounds. It should be also kept in mind that for an actual surface
the parameters may deviate from the given values due to materials science aspects not
addressed in this work.
}
\label{MaterialParameters}
\end{table}

For the numerics we normalized energies on both sides of the interface to the 
thermal energy $k_BT_e$ of the electrons emitted from the plasma source. Lengths,
in contrast, are normalized, depending on which side of the interface is considered, 
to the electron Debye length of the wall, 
\begin{eqnarray}
\lambda^w_D=\sqrt{\frac{\varepsilon k_BT_*}{8\pi n_{\rm ref}}}~,
\end{eqnarray}
or the electron Debye length of the plasma
\begin{eqnarray}
\lambda^p_D=\sqrt{\frac{k_BT_e}{8\pi n_{se}}}~. 
\end{eqnarray}
After rewriting the equations in normalized form we replace the boundary density 
ratios $n_{si}/n_{se}$, $n_{b*}/n_{\rm ref}$, and $n_{bh}/n_{\rm ref}$ by 
(\ref{Alpha}), (\ref{Beta}), and (\ref{Gamma}). We then obtain four equations
for the four (normalized) potential energy drops 
$y_1=U_1/k_BT_e$, $y_b=U_b/k_BT_e$, $y_p=U_p/k_BT_e$, and $y_w=U_w/k_BT_e$. 
They factorize into two sets of two equations each, one for $(y_p, y_w)$ and 
one for $(y_b, y_1)$. 

More specifically, the equations for $(y_p, y_w)$ arise from the 
quasi-neutrality condition (\ref{NeutralP2}) and the field-free 
condition (\ref{FieldFreeP}). After replacing the boundary ratio  
$n_{si}/n_{se}$ by (\ref{Alpha}) they become nonlinear equations 
for $(y_p, y_w)$ alone and can be casted into the form
\begin{eqnarray}
f_{\rm I}(y_p,y_w) &= 0~, \label{fI}\\
f_{\rm II}(y_p,y_w) &= 0 \label{fII}~.
\end{eqnarray}
Except for the difference arising from the different choice of the zero of 
the energy axis these two equations are identical to the ones given by
Schwager and Birdsall~\cite{SB90}. Replacing the boundary density ratios 
$n_{b*}/n_{\rm ref}$ and $n_{bh}/n_{\rm ref}$ by (\ref{Beta}) and (\ref{Gamma})
in the field-free condition (\ref{FieldFreeW}) and the recombination 
condition (\ref{ReComb}) leads to two nonlinear equations for $(y_1, y_b)$. 
They can be casted in the same form,
\begin{eqnarray}
f_{\rm III}(y_1,y_b) &= 0 \label{fIII}\\
f_{\rm IV}(y_1,y_b) &= 0 ~. \label{fIV}
\end{eqnarray}
The nonlinear equations (\ref{fI})--(\ref{fIV}) contain no additional physical
information. They are therefore not explicitly given. 

We solve (\ref{fI})--(\ref{fIV}) graphically as explained in Fig.~\ref{CondPlasmaWall} 
below. The solutions $(y_p, y_w)$ on the plasma side and $(y_1, y_b)$ on the wall side 
are linked to each other by the jump condition for the electric field (\ref{Kappa}) which 
in normalized form becomes a condition containing all four boundary density ratios. 
Since the boundary ratios $n_{b*}/n_{\rm ref}$, $n_{bh}/n_{\rm ref}$, and $n_{si}/n_{se}$ 
are known from the potential drops the fourth ratio $n_{si}/n_{\rm ref}$ can be 
determined from (\ref{Kappa}) making thereby the wall side of the double layer 
consistent with the plasma side. At this point the recombination condition (\ref{ReComb}) 
turned out to be essential. Without it the collisionless theory had not enough equations 
to determine all the unknown parameters. 

\begin{figure}[t]
\includegraphics[width=0.99\linewidth]{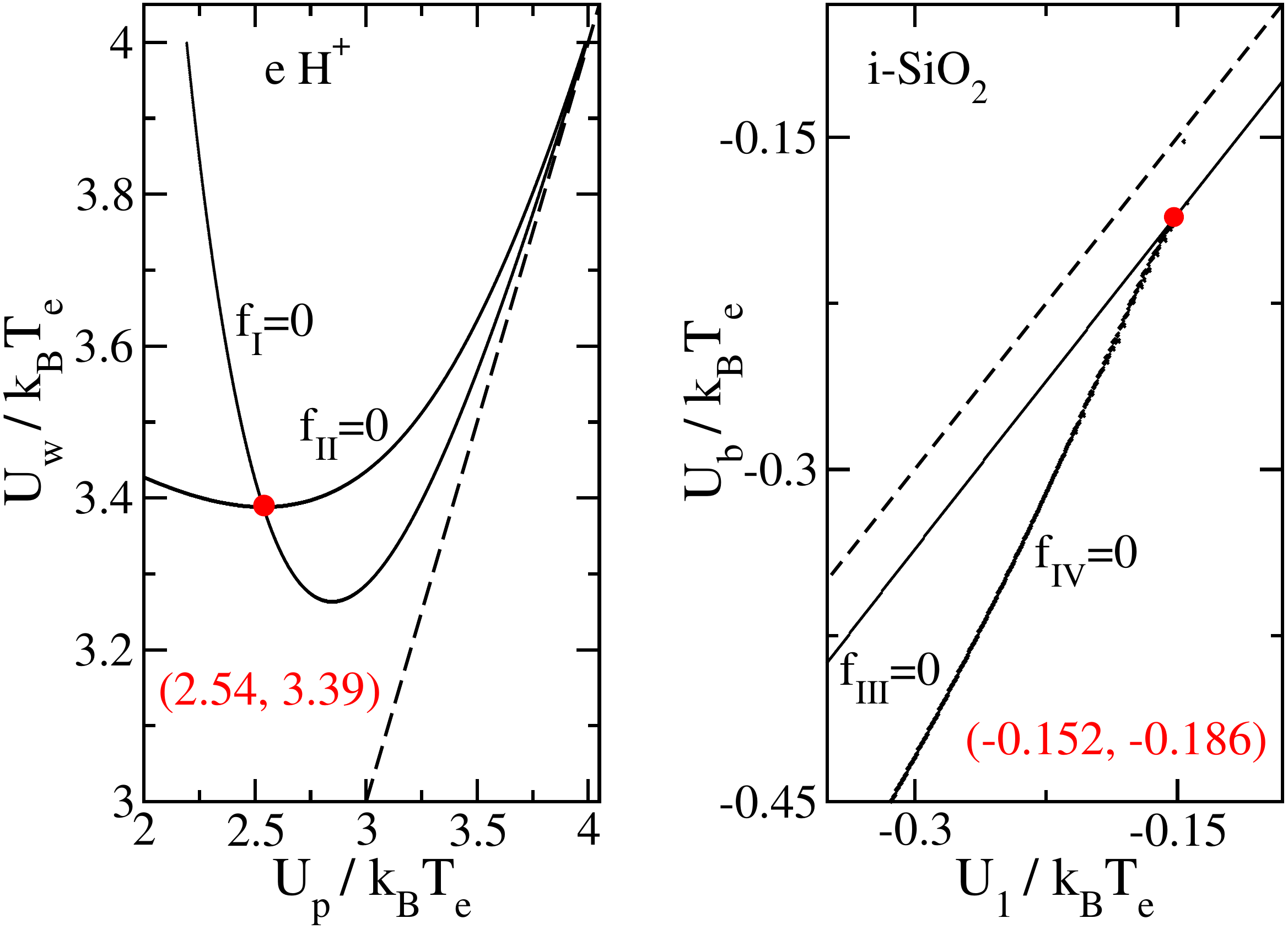}
\caption{(Color online) The left panel shows the solutions of $f_{\rm I}(y_p,y_w)=0$
and $f_{\rm II}(y_p,y_w)=0$. Where the two curves cross, indicated by the
red bullet, gives the normalized potential drops $(y_p,y_w)=(2.54,3.39)$. With
$k_BT_e=2\,{\rm eV}$ follows $U_p=5.08\,{\rm eV}$ and $U_w=6.78\,{\rm eV}$ with
$U_p$ the potential of the collector sheath. In the right
panel the graphical solution of $f_{\rm III}(y_1,y_b)=0$ and $f_{\rm IV}(y_1,y_b)=0$
is shown leading to the normalized potential drops $(y_1,y_p)=(-0.152,-0.186)$,
again marked by the red bullet. Hence, $U_1=-0.304\,{\rm eV}$ and $U_b=-0.372\,{\rm eV}$.
The potential drop $U_1$ is the band bending inside the wall. The
effective mass of the conduction band electrons is $m_*=m_e/1.3$. Plasma parameters and
the remaining wall parameters are given in Table~\ref{MaterialParameters}.}
\label{CondPlasmaWall}
\end{figure}

For the numerical calculations we take the parameters given in Table~\ref{MaterialParameters}.
A few comments about this choice are in order. The thermal energy of the electrons 
released from the  plasma source is $k_BT_e=2\,{\rm eV}$. For the ions we take 
$k_BT_i=0.2\,{\rm eV}$. The ion temperature is thus rather 
high but decreasing it further produced numerical instabilities already on 
the plasma side. Since it is reasonable to assume the ions to set the lowest thermal 
energy for the charged species, we also set 
$k_BT_*=k_BT_h=0.2\,{\rm eV}$. For the small band gap material \Si\ 
the intrinsic charge density (\ref{nint}) is then rather high, leading to 
unrealistically high densities in the double layer. But making the reservoir 
colder than the coldest species of the sources seems to be physically not plausible
and might have made the numerical solution of $f_{\rm III}=f_{\rm IV}=0$ even more 
subtle than it turned out to be already. The masses we take are the ones for a hydrogen 
plasma and a \Si\ or a \SiOTwo\ surface. In the energy range where we need the (effective)
masses of the wall's charge carriers they are a bit uncertain~\cite{vanDriel84,Evtukh01,Riffe02}. 
We used therefore the mass of the conduction band electrons as a parameter to be 
varied within physically reasonable bounds.

\begin{figure}[t]
\includegraphics[width=0.99\linewidth]{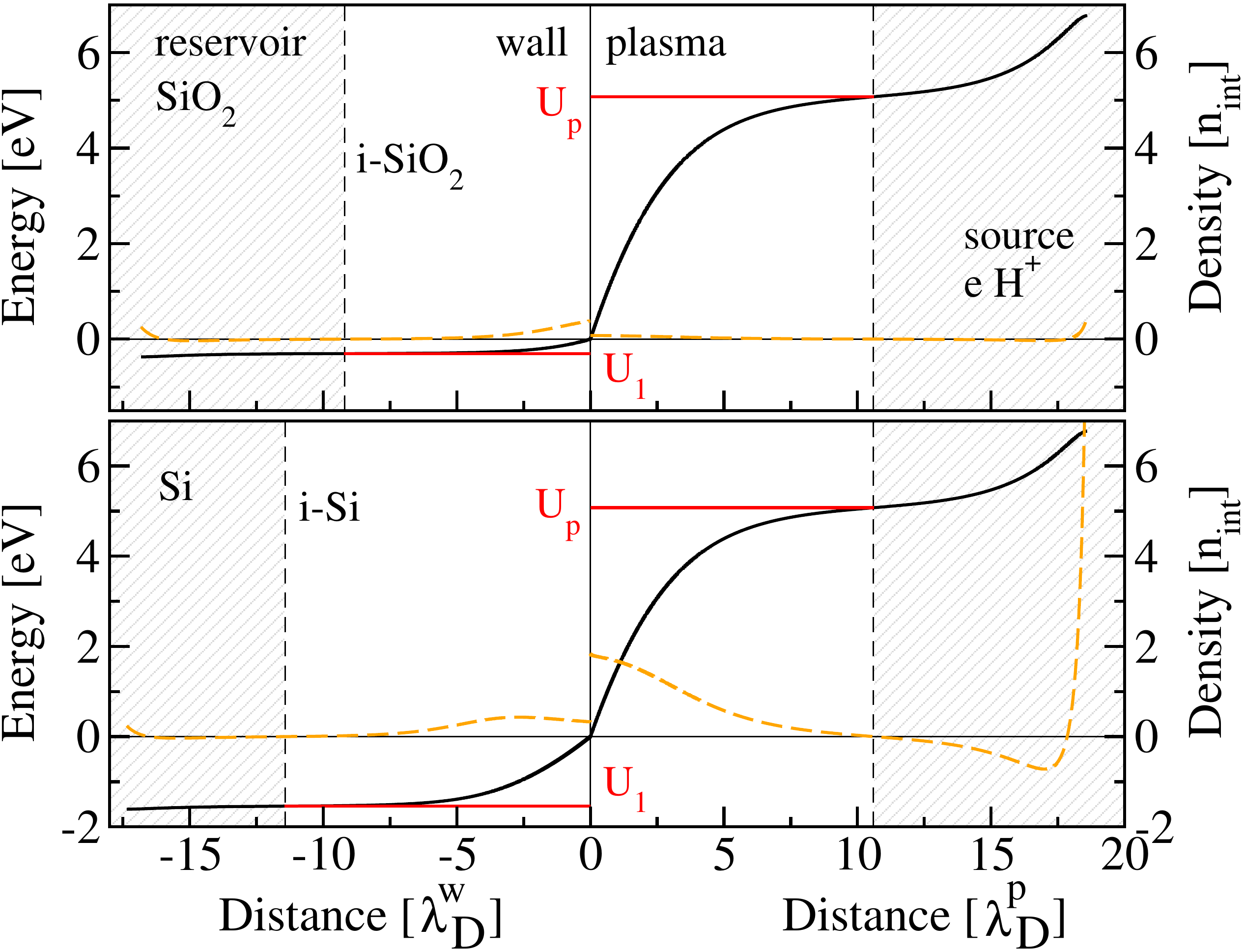}
\caption{(Color online) Potential profile for i-\SiOTwo\ (upper panel) and i-\Si\
(lower panel) in contact with a hydrogen plasma with the properties specified
in Table~\ref{MaterialParameters}. The black curves are the potential profiles
in eV (left ordinate) and the orange curves give the density profiles $\rho_{p}(z)$
and $\rho^t_{w}(z)$ as defined in (\ref{RhoPlasma1}) and (\ref{RhoWallt}) in
units of $n_{\rm int}$ (right ordinate). The grey regions indicate respectively
the reservoir and the source which have been set up to provide the correct
physical boundaries for the double layer. Relevant for the double layer is thus
only the region between the vertical dashed lines giving the position of the
inflection points $z_1$ and $z_p$ in units of the electron Debye screening lengths
$\lambda_D^w$ and $\lambda_D^p$, respectively. Hence, the sheath potential is
given by $U_p$ and the band bending by $U_1$. The mass of the conduction band
electrons is $m_*=m_e/1.3$ for \SiOTwo\ and $m_*=m_e/2$ for \Si . The
remaining surface parameters can be found in Table~\ref{MaterialParameters}
together with the plasma parameters.}
\label{CombinedPotProfile}
\end{figure}

As mentioned, to determine the simultaneous roots of $f_{\rm I}(y_p,y_w)=0$ 
and $f_{\rm II}(y_p,y_w)=0$ on the one hand and $f_{\rm III}(y_1,y_b)=0$ and  
$f_{\rm IV}(y_1,y_b)=0$ on the other we used a graphical approach. First, 
we determined from  $f_{\rm I}(y_p,y_w)=0$ and $f_{\rm II}(y_p,y_w)=0$ two 
separate relations for $y_w(y_p)$. Plotting them and looking for points where 
they cross gives the simultaneous root $(y_p,y_w)$ as shown in the left 
panel of Fig.~\ref{CondPlasmaWall}. A similar procedure for 
$f_{\rm III}(y_1,y_b)=0$ and $f_{\rm IV}(y_1,y_b)=0$, depicted in the right
panel, gives the root $(y_1,y_b)$. On the plasma side the equations turned 
out to be more handy than on the wall side as can be seen by the noisiness
of the roots of $f_{\rm IV}$. Nevertheless in all the cases we discuss in 
this work the simultaneous solutions of the nonlinear equations 
(\ref{fIII}) and (\ref{fIV}) are stable and reproducible. 
After the normalized energies $y_b$, $y_1$, $y_p$, and $y_w$ are known, the 
density ratios $n_{b*}/n_{\rm ref}$, $n_{bh}/n_{\rm ref}$, $n_{si}/n_{se}$ and 
$n_{si}/n_{\rm ref}$ can be determined. All the parameters we introduced 
in the modeling of the double layer are then fixed and the potential profile
$U_c(z)$ can be calculated by integrating (\ref{UcPlasma}) and (\ref{UcWall}).

Using $m_*=m_e/1.3$ and $m_*=m_e/2$, respectively, for the masses of the 
conduction band electrons in i-\SiOTwo\ and i-\Si , which we consider most 
reasonable for the elevated temperatures~\cite{vanDriel84,Riffe02}, we plot in 
Fig.~\ref{CombinedPotProfile} the potential profiles (black curves) together with 
the profiles for the charge density (orange curves). The density profiles show that 
the source as well as the reservoir are not charge-neutral as expected. Charge-neutrality 
is satisfied by construction only around the inflection points $z=z_1$ and $z=z_p$. 
It is interesting to note that for i-\Si\ the density of the space charge increases 
not monotonously towards the wall. We attribute this to the unrealistically high intrinsic 
charge concentration $n_{\rm int}$ arising from the high temperature. The band gap in 
\Si\ is only around $1$\,eV and thus only five times larger than the thermal energy 
of the charge carriers. The potential drops on the wall and the plasma side 
relevant for the double layer are $U_1$ and $U_p$. The former is the plasma-induced
band bending whereas the latter is the wall-induced sheath potential. The potential
drops from $z=z_w$ to $z=z_p$ and from $z=z_1$ to $z=z_b$ on the other hand are 
required to establish the Maxwellians at the boundaries.
The flatness around the inflection points depends on the accuracy with which the 
roots $(y_b, y_1)$ and $(y_p, y_w)$ have been determined. Increasing the accuracy
makes the plateaus wider but the differences $y_1-y_b$ and $y_w-y_p$ remain the same.

\begin{figure}[t]
\includegraphics[width=0.99\linewidth]{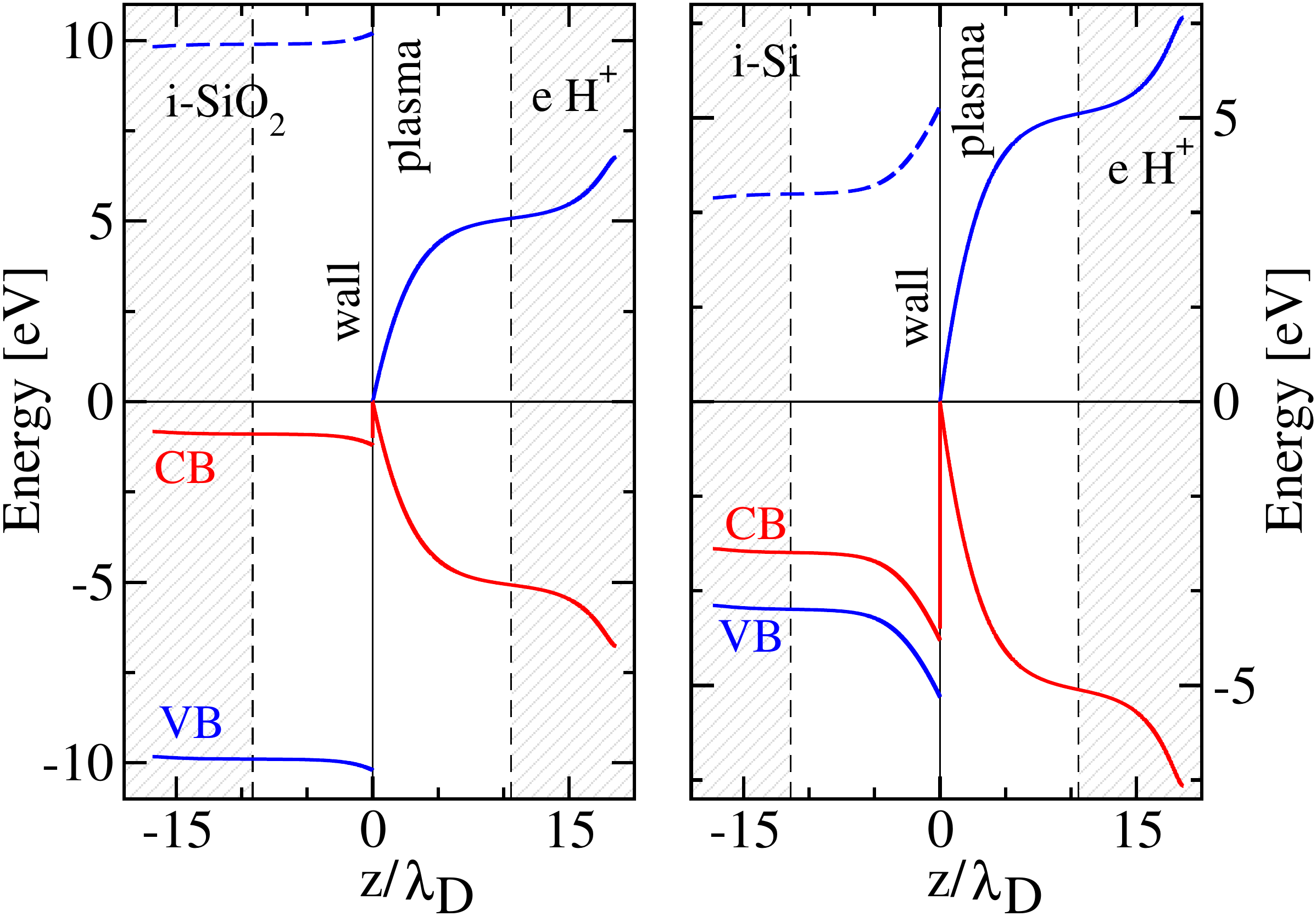}
\caption{(Color online) Band edges for i-\SiOTwo\ (left panel) and i-\Si\ (right
panel) in contact with a hydrogen plasma with the parameters given in
Table~\ref{MaterialParameters}. Inside the wall the solid red (blue) curves are
the edges for the conduction (valence) band and in front of the wall the red (blue)
curves give the potential energy for electrons (ions). The dashed blue lines are
the edges for the valence band holes. Lengths are given in units of the
electron Debye screening lengths $\lambda_D^w$ on the wall side and $\lambda_D^p$
on the plasma side. For \SiOTwo\ the effective mass for the conduction band
electrons is $m_*=m_e/1.3$ while for \Si\ it is $m_*=m_e/2$. The other parameters
can be found in Table~\ref{MaterialParameters}.}
\label{CombinedEdges}
\end{figure}

The potential profile $U_c(z)$ determines via (\ref{Uvb}) and (\ref{Ucb}) the edges 
of the bands inside the wall. For i-\SiOTwo\ and i-\Si\ this is shown in 
Fig.~\ref{CombinedEdges} together with the potential energies in front
of the wall. The mass of the conduction band electrons is again $m_*=m_e/1.3$ 
for \SiOTwo\ and $m_*=m_e/2$ for \Si . From the discussion of Fig.~\ref{CombinedPotProfile}
it is clear that in the grey regions the boundary conditions are established. Hence,
only the profiles between the grey regions apply directly to the double layer. The 
figure gives an idea how strong a hydrogen plasma with the parameters specified in 
Table~\ref{MaterialParameters} modifies the band structure of i-\SiOTwo\ and 
i-\Si . For n- and p-type \SiOTwo\ and \Si\ surfaces the band structure would 
look qualitatively similar. The absolute numbers, however, would be different because 
the reference densities $n_{\rm ref}$ are different leading for instance to  
different electron Debye screening lengths. Instead of plotting the band structures 
for different dopings we summarize representative results for doped surfaces in 
Table~\ref{TableForData}. The surfaces are always in contact with the hydrogen plasma
specified in Table~\ref{MaterialParameters}.

Besides the band bending $U_1$ and the sheath potential $U_p$ Table~\ref{TableForData}
contains also data for the net electron and ion flux $j=j_e=j_i$ towards the interface 
and the charge density $N_{\rm EDL}$ in one leg of the double layer, 
\begin{eqnarray}
N_{\rm EDL}=\int_{z_1}^0 \!\!dz \rho^t_w(z)=\int_0^{z_p}\!\!\!\!dz \rho_p(z) ~.
\end{eqnarray}
Since between $z=z_1$ and $z=z_p$ the double layer is charge neutral, the 
charge density confined inside the wall between $z=z_1$ and $z=0$ is by construction
equal to the charge density between $z=0$ and $z=z_p$ on the plasma side. Numerically
the integrated charge densities coincide better than one percent. The intrinsic
charge density $n_{\rm int}$ depends according to (\ref{nint}) on the mass of the 
conduction band electrons. Varying $m_*$ changes thus even for fixed $x=\infty$ (intrinsic), 
$x=n_{\rm int}/n_A$ (p-type) and $x=n_{\rm int}/n_D$ (n-type) the reference density 
$n_{\rm ref}=n_{\rm int}$ (intrinsic), $n_{\rm ref}=n_p$ (p-type), and  $n_{\rm ref}=n_{n}$ 
(n-type) and thus the boundary condition to be met by the double layer on the wall side. 
As a result, the properties of the double layer depend on $m_*$ as can 
be seen in Table~\ref{TableForData}.  

For \Si\ surfaces the fluxes $j$ and the charge densities $N_{\rm EDL}$ trapped
in one leg of the double layer are rather high. The reason is the high ion 
temperature required to stabilize the numerical calculations. Since the ion 
temperature sets also the scale for the temperature of the charge carriers inside 
the wall, the intrinsic density $n_{\rm int}$, and with it the reference densities 
$n_{\rm ref}$, are very high for small band gap materials such as \Si . The 
unrealistically high densities of the reservoir lead also to unrealistically high 
densities in the plasma source. As a result, the electron Debye screening lengths 
$\lambda_D^{w,p}$ are extremely short making the space charges on both sides of 
the interface very narrow, at most $100\,\AA$ wide. For any realistic gas pressure 
the double layer would be thus collisionless on the plasma side. It would be even 
almost collisionless on the solid side, because the inelastic scattering length 
$l_{\rm inel}\approx 100\, \AA$~\cite{Lueth15}. For \Si:${\rm H}^+$-e the premises of our 
numerical calculations are thus satisfied but only for a situation which in practice 
cannot be realized.

The data for \SiOTwo:${\rm H}^+$-e are more realistic because the band gap of \SiOTwo\ 
is significantly larger than the thermal energy of the reservoir leading to 
reasonable reference densities. The fluxes $j$ and the charge densities $N_{\rm EDL}$
are then also more realistic. Due to the lower densities the electron Debye screening 
lengths are much larger leading to wider space charge layers. On 
the plasma side, the space charge can still be considered collisionless because, 
based on $l_{\rm cx}=1/n_g\sigma_{\rm cx}=k_BT_g/\sigma_{\rm cx}p$ with 
$\sigma_{\rm cx}\approx 10^{-15}\,{\rm cm}^2$~\cite{LL05} the cross section for 
charge-exchange scattering and $k_BT_g=0.03\,{\rm eV}$ the gas temperature, the 
charge-exchange scattering length $l_{\rm cx}$ will be much larger than $10^{-3}\,{\rm cm}$
for any reasonable gas pressure $p$. On the solid side, however, collisions cannot be 
ignored anymore because the space charge is now much wider than the inelastic scattering 
length $l_{\rm inel}\approx 100\, \AA$~\cite{Lueth15}. It should be however recalled that the 
recombination condition (\ref{ReComb}) anticipates interband collisions. There is thus some 
hope that the data presented for \SiOTwo\ give the correct order of magnitude. But only an 
investigation which includes intra- and interband collisions inside the wall, and thus 
requires on the solid side the numerical solution of the collisional Boltzmann equations (to 
be considered beyond the scope of the present work), can tell whether this is indeed the case. 

The band bending we find for i-\SiOTwo\ in contact with a hydrogen plasma with 
$k_BT_e=2\,{\rm eV}$, $k_BT_i=0.2\,{\rm eV}$, and 
an ion density $n_{si}\approx 1.3\cdot 10^{11}\,{\rm cm}^{-3}$ (which follows from the 
numerical values for $n_{si}/n_{\rm ref}$ and $n_{\rm ref}$ given in Table~\ref{TableForData}) 
is about $0.3\,{\rm eV}$. In our previous work~\cite{HBF12} we obtained for a helium plasma 
with $k_BT_e=2\,{\rm eV}$, $k_BT_i=0$, and a plasma density $n_0=10^7\,{\rm cm}^{-3}$ 
a band bending of about $0.1\,{\rm eV}$. In view of the much denser plasma considered
in the present work the larger band bending we find now is to be expected. Thus, the numbers 
we obtain are consistent with our previous work. In contrast to it, however, the new approach
presented in this paper, is more general because it does not dependent on a thermodynamical 
principle. Working with Boltzmann equations and matching conditions for the distribution 
functions gives us more flexibility in treating ions, electrons, valence band holes, 
and conduction band electrons. In addition, it is conceivable by using time-dependent 
Boltzmann equations to study within this framework also the temporal build-up of the 
double layer. Time-dependent Boltzmann equations would be also required for tracking
the time-periodic modifications of the band structure induced by 
radio-frequency (micro)discharges. The approach we presented in this paper could be
also applied to this problem.

\section{Conclusions}
\label{Conclusions}

We presented a semiclassical kinetic theory for the quasi-stationary electric double 
layer at a planar dielectric surface in contact with a two-temperature plasma. It is 
based on the Poisson equation for the electric potential energy and two sets of 
Boltzmann equations, operating in disjunct half-spaces, for the species involved 
in the electronic response of the plasma and the solid: electrons and ions on the 
plasma side and conduction band electrons and valence band holes on the wall side. 
Crucial for the kinetic theory are the matching conditions for the distribution 
functions across the interface. For electrons they are identical to the ones 
employed in solid heterostructures whereas for the ion and hole distribution
functions they represent a hole injection model. As boundary conditions far 
away from the interface we use Maxwellian distribution functions. The species'
boundary densities as well as the potential drops in the double layer are determined 
in the course of the calculation which at the end yields self-consistent potential 
and density profiles across the interface. 

\begin{figure}[t]
\begin{center}
\includegraphics[height=0.3\linewidth]{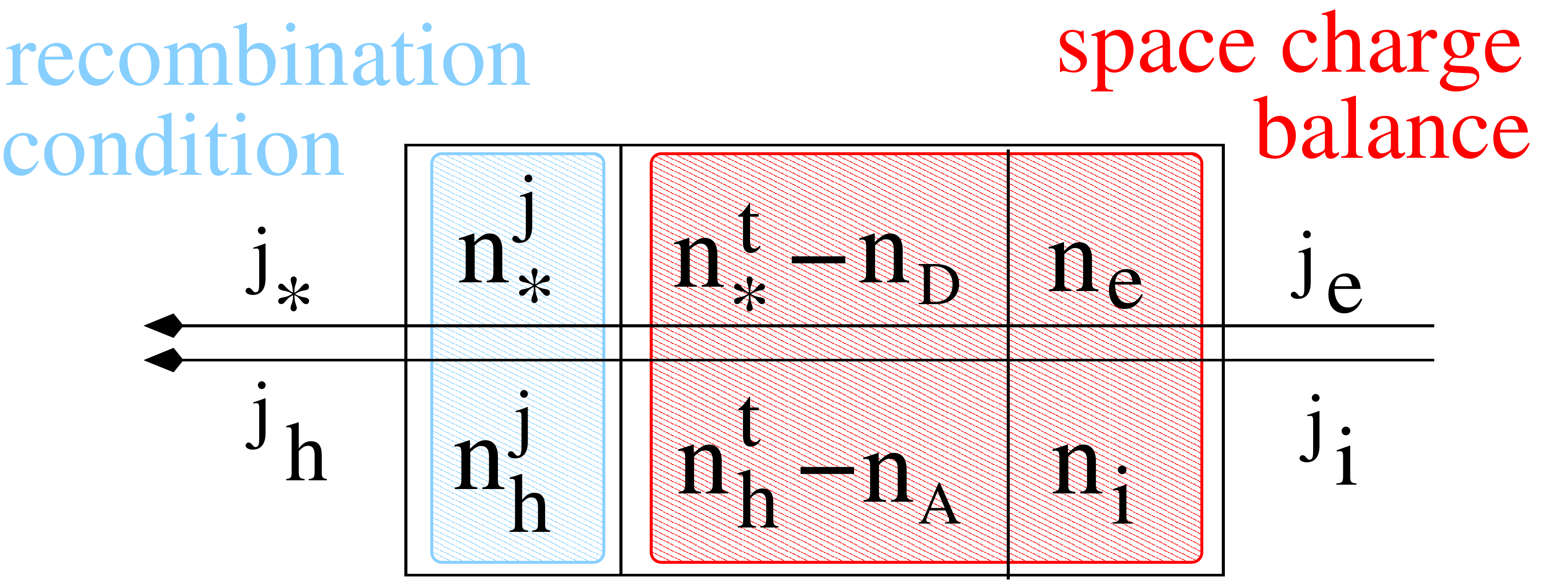}
\end{center}
\caption{(Color online) Quintessence of our kinetic approach for modeling a quasi-stationary
         electric double layer at a plasma-facing solid. The net positive space charge in the 
         plasma sheath $\rho_p(z)=n_i(z)-n_e(z)$ is balanced by the thermalized/trapped net 
         negative space charge inside the wall $\rho_w^t(z)=n_*^t(z)-n_D-n_h^t(z)+n_A$. The 
         non-thermalized charges $\rho_w^j(z)=n_*^j(z)-n_h^j(z)$ inside the wall, arising from 
         the continuing influx of electrons and ions recombine and limit thereby in conjunction 
         with the matching and boundary conditions the fluxes 
         $j_e$ and $j_i$ released from the plasma source. By extending the modeling into the 
         wall and distinguishing in it thermalized/trapped from non-thermalized charge carriers 
         the equality of electron and ion fluxes can be reconciled with a vanishing ion density 
         at the interface because not the ions matter directly but the holes in the valence 
         band.}
\label{Species}
\end{figure}

The physical picture implemented in our approach is that of a quasi-stationary electric 
double layer sandwiched between quasi-neutral, field-free regions. On the plasma side 
this region mimics the bulk plasma whereas on the wall side it mimics either an intrinsic 
or an extrinsic dielectric wall. To enforce these physical boundary conditions a reservoir
for conduction band electrons and valence band holes has to be attached to the wall side
while on the plasma side an electron and ion source is required. The reservoir and the 
source per se have no direct physical meaning. They are only technical devices to implement 
the physical boundary conditions. 

Figure~\ref{Species} summarizes the essence of our approach. The net electron and ion fluxes 
originating from the plasma source lead to a double layer consisting of electrons and ions 
on the plasma side and thermalized as well as non-thermalized conduction band electrons and 
valence band holes on the wall side. The net positive space charge of the double layer on the 
plasma side is balanced inside the wall only by the thermalized net negative space charge. 
Recombination of the wall's non-thermalized charge densities (which in the collisionless 
theory is taken into account by an ad-hoc recombination condition) limits in conjunction 
with the matching and boundary conditions the net electron 
and ion influx after the double layer is established. Inside the wall the ion flux becomes 
a flux of valence band holes, whereas the flux of electrons becomes a flux of conduction band 
electrons. At quasi-stationarity these two fluxes have to give rise to a permanent flux of 
photons and/or phonons depending on whether the recombination occurs radiatively or 
non-radiatively. This is however not yet included in the modeling.

Applying the kinetic approach to a collisionless, perfectly absorbing plasma-wall 
interface with material parameters representing \Si\ and \SiOTwo\ surfaces in contact 
with a hydrogen plasma demonstrated its feasibility. Since the wall sets an absolute 
scale for energy and density we could determine absolute numbers for the potential 
drops, the charge densities on either side of the double layer, and the particle 
fluxes maintaining the double layer. In view of the high ion temperature necessary to 
stabilize the numerics on the plasma side, which in turn implied a high temperature 
for the charge carriers inside the wall, the numerical data for the small band gap 
material \Si\ turned out to be unrealistic. For the large band gap material 
\SiOTwo , however, we obtained rather reasonable results showing that the kinetic
approach presented in this work has the potential to describe the electric response 
of a real surface to a real plasma and vice versa if it is applied to a more realistic 
model of the plasma-wall interface. Not only collisions and quantum-mechanical reflection
from the interface have then to be included but also a realistic description
of the wall's band structure. 

Obviously for a realistic interface model the kinetic equations become rather complex, 
even in the quasi-stationary regime. Iterative solution strategies have then to go
hand-in-hand with Monte-Carlo techniques to tackle the problem. It may be also required
to reduce the complexity by a systematic multiple-time scale analysis, anticipating
the different time scales on which intra- and interband scattering processes occur, 
which we expect to separate the loss processes due the recombination of conduction band
electrons with valence band holes from the thermalizing collisions. The ad-hoc 
recombination condition we employed in the modeling of the collisionless, perfectly 
absorbing interface may then turn out to be a secular condition in a multiple time-scale
analysis. We leave this conjecture for future investigations. 

No matter how powerful the techniques for attacking the kinetic equations, the quality
of the results depends on the quality of the electronic band structure. Even in the 
absence of the uncertainties arising from plasma-induced imperfections and chemical
contaminations, getting all the required band structure information is a challenge.
Since electrons and holes are injected into states far away from the band edges, knowing 
the band structure in the vicinity of the band gap is not sufficient. The dispersion 
of the conduction (valence) band has to be known all the way up (down) to the 
electron affinity (ionization level of the ions in the plasma). To restrict the
considerations to two bands with one energy valley may in fact be unjustified 
as well as the effective mass approximation. The kinetic modeling of the electronic 
response of a realistic plasma-wall interface has thus to go hand in hand with advanced
surface diagnostics and ab-initio band structure calculations. 

\ack
Support from the Deutsche Forschungsgemeinschaft through Project B10
of the Transregional Collaborative Research Center SFB/TRR 24 is greatly
acknowledged.\\

\appendix 

\section{Mathematical technicalities} 
\label{Functions}

In this appendix we give for the collisionless, perfectly absorbing plasma-wall interface
the formulae for the charge densities $n_s(U_c)$ with $s=e,i,*,h$ and the integrated
charge densities ${\cal F}_{e,i}(U_c)$ and ${\cal G}^t_{*,h}(U_c)$
from which
   \begin{eqnarray}
       {\cal F}(U_c) &=\sqrt{[{\cal F}_i(U_c)]^2-[{\cal F}_e(U_c)]^2}~, \\
       {\cal G}^t(U_c) &= \sqrt{[{\cal G}_*^t(U_c)]^2-[{\cal G}_h^t(U_c)]^2}
   \end{eqnarray}
follow. These functions can 
be obtained, as described in Sect.~\ref{Formalism}, from the solutions of the collisionless 
Boltzmann equations (\ref{BTECollLess}) constructed from the trajectory analysis explained 
in Fig.~\ref{TA}. 

On the plasma side the charge densities are
\begin{eqnarray}
n_i(U_c) &=\! \frac{n_{si}}{2}\exp\bigg(\frac{U_w-U_c}{k_BT_i} \bigg)
           \Phi_c\bigg(\sqrt{\frac{U_w-U_c}{k_BT_i}}\bigg) ~, \label{ni}\\
n_e(U_c) &=\! \frac{n_{se}}{2}\exp\bigg(\frac{U_c-U_w}{k_BT_e} \bigg)
          \bigg[1 + \Phi\bigg(\sqrt{\frac{U_c}{k_BT_e}}\bigg)\bigg] ~.
\label{ne}
\end{eqnarray}

On the wall side we distinguish thermalized from non-thermalized charge carriers. The 
density of conduction band electrons is therefore given by  
\begin{eqnarray}
n_*(U_c)=n_*^t(U_c) + n_*^j(U_c)
\label{n*}
\end{eqnarray}
with 
\begin{eqnarray}
n_*^t(U_c) &= n_{b*} \exp\bigg(\frac{U_c-U_b}{k_BT_*}\bigg)\bigg[
             \Phi\bigg(\sqrt{\frac{U_c+\chi}{k_BT_*}}\bigg) \nonumber\\
           &-  \Phi\bigg(\sqrt{\frac{U_c-U_b}{k_BT_*}}\bigg) \bigg]
\label{n*t}
\end{eqnarray}
the density of the thermalized conduction band electrons and 
\begin{eqnarray}
n_*^j(U_c) &= n_{si}\bigg(\frac{m_*}{4m_i}\frac{k_BT_i}{k_BT_e}\bigg)^{1/2}
             \exp\bigg(\frac{\chi+U_c}{k_BT_e}\bigg) \nonumber\\
           &\times  \Phi_c\bigg(\sqrt{\frac{\chi+U_c}{k_BT_e}}\bigg)
\label{n*j}
\end{eqnarray}
the density of the non-thermalized conduction band electrons. 

Similarly, the density of valence band holes is written as 
\begin{eqnarray}
n_h(U_c)=n_h^t(U_c) + n_h^j(U_c)
\label{nh}
\end{eqnarray}
with 
\begin{eqnarray}
n_h^t(U_c) = n_{bh} \exp\bigg(\frac{U_b-U_c}{k_BT_h}\bigg)
\label{nht}
\end{eqnarray}
the thermalized density. The non-thermalized density reads 
\begin{eqnarray}
n_h^j(U_c) = n_{si} \bigg(\frac{m_h}{m_i}\frac{k_B T_i}{\pi} \bigg)^{1/2} \frac{\Psi(U_c)}{2\Gamma}
\label{nhj}
\end{eqnarray}
with 
\begin{eqnarray}
\Psi(U_c)=2(\sqrt{A-U_c}-\sqrt{B-U_c})
\label{Psi}
\end{eqnarray}
and $A=I+(\Gamma/2)-E_g-\chi$ and $B=I-(\Gamma/2)-E_g-\chi$. The particular form of $\Psi(U_c)$ 
is the result of the uniform injection probability employed in the hole injection model.

\vspace{4mm}
\noindent
The integrated densities, the Sagdeev-type potentials~\cite{SB83,Raadu89,Charles07} in our double 
layer theory, are 
\begin{eqnarray}
[{\cal F}_i(U_c)]^2 &= 8 \pi n_{si}k_BT_i \bigg[\exp\bigg(\frac{U_w-U_c}{k_BT_i}\bigg) -1 \nonumber\\
                    &- H\bigg(\frac{U_w-U_c}{k_BT_i},0 \bigg) \bigg] 
\label{Fi}
\end{eqnarray}
for the ions, 
\begin{eqnarray}
[{\cal F}_e(U_c)]^2 &= 8\pi n_{se}k_BT_e \bigg[1 - \exp\bigg(\frac{U_c-U_w}{k_BT_e}\bigg)   \nonumber\\
                    &+ \exp\bigg(-\frac{U_w}{k_BT_e}\bigg) H\bigg(\frac{U_w}{k_BT_e},\frac{U_c}{k_BT_e} \bigg) \bigg] 
\label{Fe}
\end{eqnarray}
for the electrons, 
\begin{eqnarray}
[{\cal G}^t_*(U_c)]^2 &=\frac{16\pi}{\varepsilon}n_{*b}k_BT_*\bigg[ 
                       \exp\bigg(-\frac{U_b+\chi}{k_BT_*}\bigg) \nonumber\\
                      &\times H\bigg(\frac{U_c+\chi}{k_BT_*},\frac{U_b+\chi}{k_BT_*}\bigg)  
                      -H\bigg(\frac{U_c-U_b}{k_BT_*},0\bigg) \bigg] \nonumber\\
                     &- \frac{16\pi}{\varepsilon} n_D (U_c-U_b)
\label{G*t}
\end{eqnarray}
for thermalized conduction band electrons and 
\begin{eqnarray}
[{\cal G}^t_h(U_c)]^2 &= \frac{16\pi}{\varepsilon}n_{hb}k_BT_h\bigg[ 1 
                       - \exp\bigg(\frac{U_b-U_c}{k_BT_h}\bigg)\bigg] \nonumber\\
                      &-  \frac{16\pi}{\varepsilon} n_A (U_c-U_b)
\label{Ght}
\end{eqnarray}
for thermalized valence band holes.

The function 
\begin{eqnarray}
H(a,b)=\int_a^b dy e^y\Phi(\sqrt{y}) &= e^b\Phi(\sqrt{b}) - e^a\Phi(\sqrt{a}) \nonumber\\
                                     &- \sqrt{\frac{4b}{\pi}} + \sqrt{\frac{4a}{\pi}}
\label{Hfct}
\end{eqnarray}
is an auxiliary function and 
\begin{eqnarray}
\Phi(\sqrt{y}) = \frac{1}{\sqrt{\pi}} \int_0^y dx \frac{e^{-x}}{\sqrt{x}}
\label{Errfct}
\end{eqnarray}
is the error function. As usual the function $\Phi_c(\sqrt{y})$ appearing in 
(\ref{ni}) is the complementary error function. 

\section*{References}

\providecommand{\newblock}{}

\vspace{350mm}

\noindent

\newpage 

\begin{table}[t]
  \begin{tabular}{|l||lllll||lll||llll|}
    \hline
    Wall      & x        & $\frac{m_e}{m_*}$  & $n_{\rm ref}$         & $\lambda_D^w$  & $\lambda_D^p$ & $\frac{n_{b*}}{n_{\rm ref}}$ & $\frac{n_{bh}}{n_{\rm ref}}$ & $\frac{n_{si}}{n_{\rm ref}}$ & $U_1$ & $U_b$ & $N_{\rm EDL}$     & $j$                            \\
              &          &                    & ($10^{18}$            & ($10^{-8}$     & ($10^{-8}$    &                              &                              &                              & (eV)  & (eV)  & ($10^{12}$        & ($10^{24}$                     \\
              &          &                    &  ${\rm cm}^{-3})$     & ${\rm cm})$    & ${\rm cm})$   &                              &                              &                              &       &       & ${\rm cm}^{-2}$)  & ${\rm s}^{-1}{\rm cm}^{-2}$)   \\\hline
    p-\Si     & 0.6      & 2.0                &  56.28                & 15.35          & 3.36          & 0.26                         & 1.03                         & 80.11                        & -1.46 & -1.47 & 28.21             & 129.14                         \\
              &          & 1.5                &  70.02                & 13.78          & 6.5           & 0.26                         & 1.03                         & 17.23                        & -0.45 & -0.46 & 14.5              & 33.34                          \\\hline
    i-\Si     & $\infty$ & 2.0                &  26.32                & 22.43          & 6.69          & 1.75                         & 1.42                         & 43                           & -1.59 & -1.66 & 14.28             & 31.7                           \\
              &          & 1.5                &  32.73                & 20.14          & 11.61         & 1.75                         & 1.42                         & 11.52                        & -0.5  & -0.57 & 8.32              & 10.43                          \\\hline
    n-\Si     & 1.7      & 2.0                &  35.09                & 19.40          & 18.18         & 2.12                         & 1.16                         & 4.36                         & -1.76 & -1.91 & 5.0               & 4.23                           \\
              &          & 1.5                &  43.73                & 17.42          & 17.58         & 2.13                         & 1.17                         & 3.76                         & -0.52 & -0.67 & 5.61              & 4.54                           \\\hline
              &          &                    &                       &                &               &                              &                              &                              &       &       &                   &                                \\
              &          &                    & ($10^{10}$            & ($10^{-3}$     & ($10^{-3}$    &                              &                              &                              & (eV)  &  (eV) & ($10^{7}$         & ($10^{15}$                     \\
              &          &                    &  ${\rm cm}^{-3}$)     & ${\rm cm}$)    & ${\rm cm}$)   &                              &                              &                              &       &       &  ${\rm cm}^{-2}$) & ${\rm s}^{-1}{\rm cm}^{-2}$)   \\\hline
    p-\SiOTwo & 0.6      & 1.3                &  12.15                & 1.91           & 4.15          & 0.51                         & 1.08                         & 2.44                         & -0.29 & -0.30 & 22.96             & 8.22                           \\
    i-\SiOTwo & $\infty$ &                    &  7.49                 & 2.43           & 6.21          & 1.76                         & 1.42                         & 1.76                         & -0.30 & -0.37 & 15.07             & 3.66                           \\
    n-\SiOTwo & 1.7      &                    &  9.99                 & 2.1            & 8.7           & 2.16                         & 1.18                         & 0.67                         & -0.32 & -0.47 & 11.21             & 1.85                           \\\hline
  \end{tabular}
\caption{Numerical data for intrinsic and extrinsic \SiOTwo\ and \Si\ surfaces exposed to the hydrogen
plasma specified in Table~\ref{MaterialParameters}.}
\label{TableForData}
\end{table}

\end{document}